\documentclass[journal]{IEEEtran}

\usepackage{cite}
\usepackage{amsmath,amssymb,amsfonts}
\usepackage{algorithmic}
\usepackage{graphicx}
\usepackage{textcomp}
\usepackage{xcolor}
\usepackage{cuted}
\usepackage{epstopdf}
\usepackage{stfloats}
\usepackage{algorithm}
\usepackage{enumerate}
\usepackage{caption}
\usepackage{comment}

\usepackage[caption=false,font=footnotesize,labelfont=rm,textfont=rm]{subfig}

\usepackage{svg}

\usepackage{hyperref}
\usepackage{amsthm}
\theoremstyle{plain}

\begin{document}

%Throughput Fairness for Movable Intelligent Surface–Enabled Periodic IoT Networks: Static Phase-Shift Design and User Scheduling
%Movable Intelligent Surface (MIS)-Enabled Wireless Communications: When Static Phase Element Meets Tunable Relative Position
\title{Movable Intelligent Surface-Enabled Wireless Communications: Static Phase Shifts with Mechanical Reconfigurability}
\author{Ziyuan~Zheng, Qingqing~Wu, Wen~Chen, Weiren~Zhu, and Ying~Gao  \vspace{-18pt} 
\thanks{The authors are with the Department of Electronic Engineering, Shanghai Jiao Tong University, 200240, China (e-mail: \{zhengziyuan2024, qingqingwu, wenchen, weiren.zhu, yinggao\}@sjtu.edu.cn).}
}% <-this % stops a space

\markboth{}%
{Shell \MakeLowercase{\textit{et al.}}: Bare Demo of IEEEtran.cls for IEEE Journals}

\maketitle


\begin{abstract}
Intelligent surfaces that reshape electromagnetic waves are regarded as disruptive technologies for wireless networks. However, existing designs sit at two costly extremes: dynamic reconfigurable intelligent surfaces (RISs) offer fine beam control but require dense cabling, continuous power consumption, and substantial signaling overhead, whereas low-cost static surfaces require no control lines or electronics but are limited to a single beam pattern. This disparity leaves a practical gap for quasi-static environments, such as industrial Internet-of-things and smart agriculture scenarios, where channels are stable with user demands changing only occasionally or periodically, and neither extreme is sufficiently economical or flexible.
To bridge this gap, we propose a novel movable intelligent surface (MIS) architecture, whose beam patterns are switched not by electronic phase tuning but by mechanically sliding a small, pre-phased secondary metasurface layer across a larger, likewise static primary layer. We develop an MIS signal model that characterizes the interaction between static phase elements with dynamic geometry via binary selection matrices. Based on this model, we formulate a new type of optimization problems that jointly design static phase shifts and the overlapping position selection of MS2 (equal to beam pattern scheduling). Efficient algorithms based on the penalty method, block coordinate descent, and Riemannian manifold optimization are proposed to tackle these mixed-integer non-convex problems.
Simulation results demonstrate that the proposed MIS architecture substantially narrows the performance gap between single-layer static surfaces and dynamic RISs, providing a practical and flexible solution for quasi-static wireless applications.
\end{abstract}

% Note that keywords are not normally used for peer-reviewed papers.

% \fi
%
% For peer review papers, this IEEEtran command inserts a page break and
\IEEEpeerreviewmaketitle

\vspace{-12pt}
\section{Introduction}
Wireless communication systems are rapidly evolving towards the sixth generation, driven by growing demands for higher data rates, expanded coverage, and improved spectral efficiency [1],[2]. In this context, reconfigurable intelligent surface (RIS), capable of dynamically reshaping wireless propagation environments through electronically controlled phase shifts, has emerged as a transformative technology [3],[4]. Fully dynamic RIS can flexibly implement passive beamforming to enhance coverage and capacity, greatly benefiting various wireless systems, as extensively demonstrated in the literature [5]-[10]. Although the fine-grained tunability of dynamic RIS does not require additional power amplification or radio frequency chains, it does come at the cost of considerable hardware complexity, significant power consumption, extensive cabling, and high control overhead for frequent phase adjustments [11],[12]. These practical constraints pose substantial challenges for the large-scale deployment of dynamic RISs, highlighting the trade-off between performance and system cost [13],[14]. In contrast, static surfaces, utilizing pre-designed fixed phase profiles, offer a low-cost and energy-efficient alternative by eliminating continuous phase tuning on massive reflective or transmissive elements [15]-[19]. 

Static surfaces have been effectively deployed in specific scenarios, such as aerial platforms or buildings, to extend terrestrial coverage via fixed beam patterns or beam flattening designs [16],[17]. Such implementations can establish virtual line-of-sight (LoS) links and expand coverage in static environments, instead of being applied in highly dynamic networks, preserving key performance benefits of dynamic RISs in a low-cost manner. However, the fundamental limitation of static surfaces is the confinement to a single beam pattern, or in other words, the inability to adapt beam patterns post-deployment, rendering them even unsuitable for scenarios where user locations or traffic demand change periodically [18]. This limitation is particularly problematic in semi- or quasi-static wireless environments, such as industrial Internet-of-things (IoT), automated warehouses, smart agriculture, smart homes, smart parking lots, utility infrastructure monitoring, and sensor networks, where conditions are relatively stable. However, occasional reconfiguration is still needed to accommodate periodic shifts in user distribution or network demand [20]-[23]. For these use cases, dynamic RISs may be excessive or not cost-effective, while purely static surfaces lack sufficient adaptability [24]-[28]. Therefore, bridging the flexibility–practicality gap between fully dynamic RIS and static surfaces is an essential challenge for next-generation quasi-static wireless applications.

To bridge this gap, we propose the Movable Intelligent Surface (MIS), a novel concept that integrates static phase shifts with mechanical reconfigurability, enabling dynamic beamforming without requiring electronic phase tuning. 
The MIS comprises two closely stacked metasurfaces (MSs): a larger, fixed primary layer (MS1) and a smaller, movable secondary layer (MS2), both employing pre-designed static phase profiles. 
This architecture significantly reduces the hardware costs associated with electronically phase-adjustable components, such as varactors, PIN diodes, cabling, and control circuitry. 
Instead, beamforming flexibility is achieved by mechanically repositioning MS2 relative to MS1, which alters the overlap pattern between their elements and thereby synthesizes distinct beam patterns. 
This relative displacement, referred to as the differential position shifting mechanism [26], introduces a novel dimension in system design. Compared to fully dynamic RISs, it reduces hardware and control overhead while preserving the favored beamforming adaptability absent in single-layer static surfaces. 
The MIS architecture is particularly suited for quasi-static scenarios, where the users can be efficiently served by switching among a limited set of pre-configured beam patterns via simple repositioning of MS2.

The main contributions are summarized as follows.
\vspace{-3pt}
\begin{itemize}
\item We propose a two-layer transmissive MIS architecture, where a smaller movable MS2 overlays a larger fixed MS1, enabling beam reconfiguration via mutual displacement while maintaining static phase shifts. We develop an MIS signal model that characterizes the interaction between static phase elements with dynamic geometry, using binary selection matrices and padding vectors. Based on this model, we formulate a type of optimization problems that jointly design static phase shifts and the overlapping position selection of MS2 (equal to beam pattern scheduling). These formulations encompass the minimum user data rate and total throughput maximization, and an extension to the element-wise mobility of MS2, which are intrinsically mixed-integer and non-convex, leveraging statistical channel state information (CSI) under spatially correlated Rician fading.  
\item We develop a suite of efficient algorithms for each optimization problem. Specifically, we first propose a penalty-assisted block coordinate descent (BCD) algorithm for the minimum user data rate maximization,  introducing a quadratic penalty to handle binary scheduling constraints, and employing successive convex approximation (SCA) to convexify and solve each subproblem iteratively; then, we develop a Riemannian manifold optimization algorithm to maximize the total throughput, optimizing all variables within a unified product manifold, yielding improved computational efficiency without binary or non-convex constraint relaxation or BCD procedures; we further propose a a penalty-assisted manifold optimization algorithm for the generalized element-wise movable MS2 configuration, enabling user-specific overlapping patterns of two layers, which quantifies the theoretical performance limits of the MIS architecture.
\item Extensive simulations validate the proposed MIS designs, demonstrating substantial beamforming gains for improved user fairness and throughput over single-layer static surface benchmarks. Results highlight that even a modest movable layer effectively enhances performance, bridging the gap between single-layer static surfaces and dynamic RISs. Moreover, the element-wise movable MS2 approach further achieves approximately 60\% of the performance gap between block-level MIS and dynamic RIS, underscoring its potential in quasi-static networks as a practical solution with balanced hardware complexity.
\end{itemize}
\vspace{-3pt}
The remainder of this paper is organized as follows. Section II describes the MIS communication model with its architecture and signal model. Section III formulates the optimization problems for joint static phase shift design and beam pattern scheduling. Section IV proposes an algorithm based on SCA and BCD for minimum user data rate maximization. Section V proposes a manifold optimization algorithm for total throughput maximization. Section VI proposes the penalty-assisted manifold optimization algorithm to design the optimal MS2 position with element-wise mobility. Section VII provides the experimental and numerical results. Finally, Section VIII concludes the paper.

\vspace{-6pt}
\section{MIS Communication Model}

As illustrated in Fig. 1, we consider an MIS-aided wireless relay system, where a transmissive MIS is deployed to assist an $L$-antenna base station (BS) in extending communication coverage to target users. The direct link from the BS to the users is assumed to be blocked. A virtual LoS path is established via the MIS, which refracts signals from the BS towards the users. Below, we detail the MIS architecture, its dynamic beamforming mechanism, and the MIS signal model.

\vspace{-6pt}
\subsection{MIS Architecture and Beamforming}
\setlength{\abovecaptionskip}{0pt}
\begin{figure}[t]
    \centering
    \includegraphics[width=3.1in]{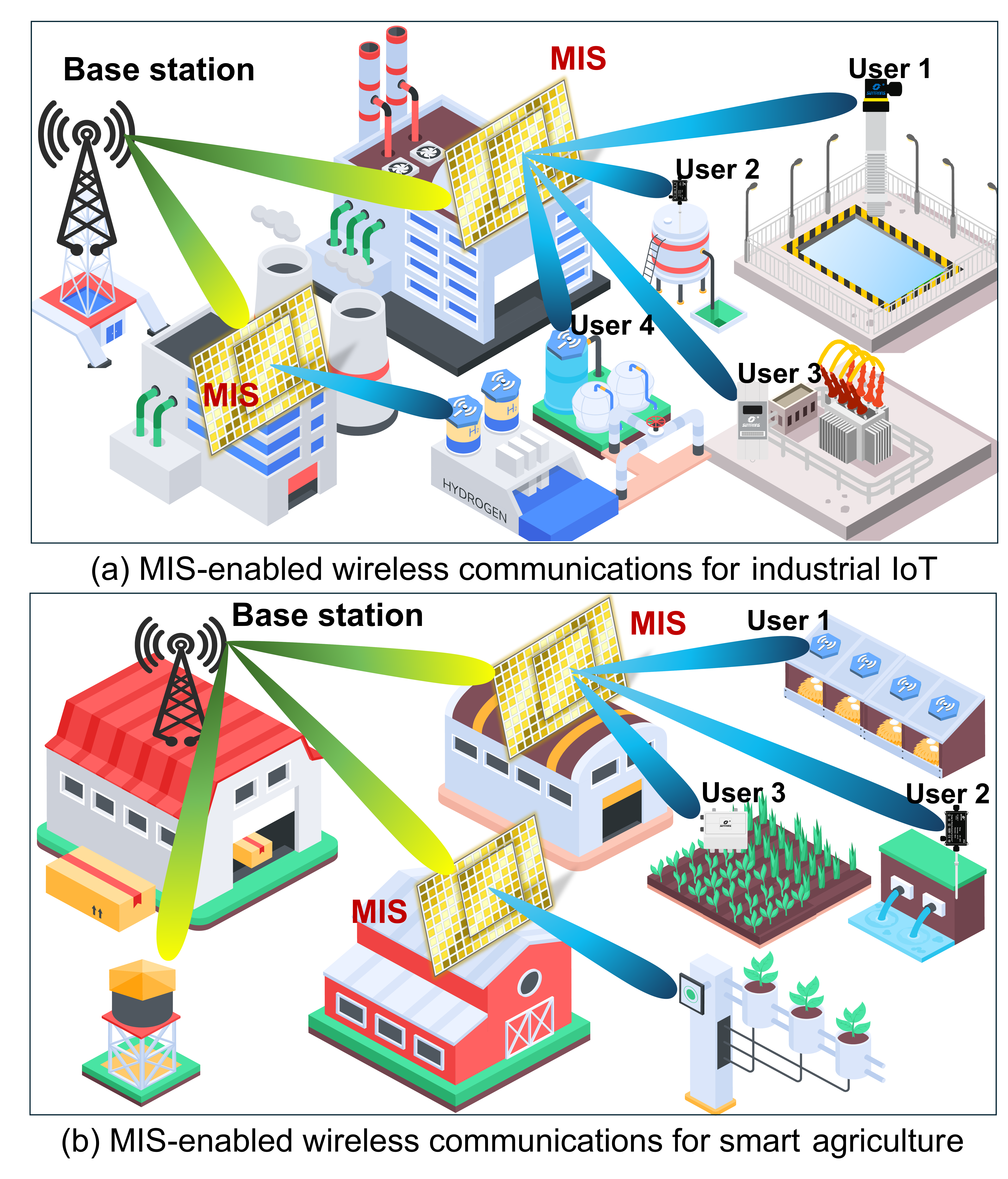}
    \captionsetup{font=small}
    \caption{Illustration of MIS-aided quasi-static wireless communication scenarios.} 
    \label{fig:system_model}
    \vspace{-6pt}
\end{figure}

\begin{figure}[t]
    \centering
    \includegraphics[width=3.2in]{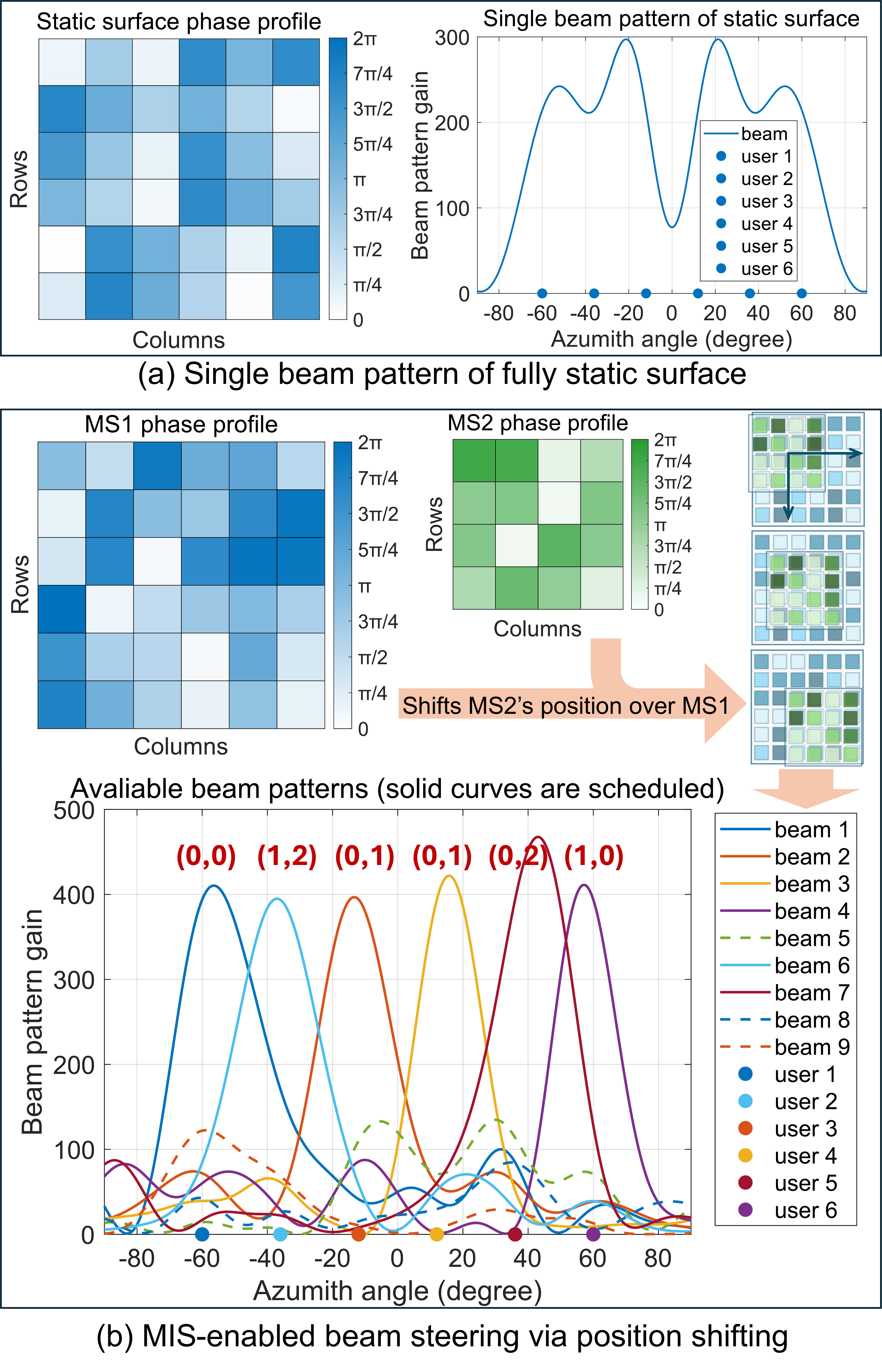}
    \captionsetup{font=small}
    \caption{An example of the MIS beam pattern switching mechanism under a max-min user signal-to-noise (SNR) rule: (a) the phase profile of a single-layer static surface and its resulting single beam pattern; (b) by sliding a small pre-phased MS2 over a larger static MS1, the MIS creates multiple composite phase maps, and hence multiple beam patterns, without electronic tuning. All possible beams are plotted, and the solid curves show those selected (the red number pairs represent relative discrete movements between MSs) for six users with significantly improved beam gain compared to case (a).} 
    \label{fig:system_model}
    \vspace{-6pt}
\end{figure}

The proposed MIS comprises two closely stacked transmissive MSs: a larger fixed-position MS1 and a smaller movable MS2. Both MS1 and MS2 employ pre-designed static phase shifts tailored to specific communication demands. MS1 consists of $M_r \times M_c$ elements arranged as a uniform planar array (UPA) with element spacing $d_{\text{MIS}}$, resulting in a total of $M = M_r \times M_c$ elements. The index of elements in MS1 follows $m = (m_r - 1) M_c + m_c$, where $m_r \in \{1,\dots,M_r\}$ and $m_c \in \{1,\dots,M_c\}$. Similarly, MS2 contains $N = N_r \times N_c$ elements, indexed as $n = (n_r - 1) N_c + n_c$, where $n_r \in \{1,\dots,N_r\}$ and $n_c \in \{1,\dots,N_c\}$. MS2 is capable of discrete movement within the boundary of MS1 along rows or columns, enabling different subsets of MS1 elements to be overlapped by MS2.
The static phase shifts of MS1 and MS2 are represented by diagonal matrices $\boldsymbol{\varPhi}=\text{diag}(\boldsymbol{\phi})\in\mathbb{C}^{M\times M}$ and $\boldsymbol{\varTheta}=\text{diag}(\boldsymbol{\theta})\in\mathbb{C}^{N\times N}$, respectively, with $\boldsymbol{\phi}=[e^{j\phi_1},\dots,e^{j\phi_M}]^T$ and $\boldsymbol{\theta}=[e^{j\theta_1},\dots,e^{j\theta_N}]^T$, where $\phi_m,\theta_n\in[0,2\pi]$ denote the static phase shifts of each element.

Unlike conventional RISs requiring electronically element-wise phase adjustments, the proposed MIS achieves dynamic beam pattern switching solely through discrete position shifts of MS2 relative to MS1. This relative movement, termed differentially position shifting, aligns the elements of MS2 with different subsets of MS1, such that the pre-designed static phases of both layers collectively synthesize distinct beam patterns without dynamic phase tuning. Fig. 2 illustrates the basic operation of the MIS and its advantage over a conventional single‐layer static surface.
The number of available MIS beam patterns is $U = U_r \times U_c$, where $U_r = M_r - N_r + 1$ and $U_c = M_c - N_c + 1$ represent the discrete shifting positions along the row and column dimensions, respectively. Each beam pattern $u\in\mathcal{U}$  is uniquely determined by an overlap position $(u_r,u_c)$, with $u = (u_r - 1) U_c + u_c$, where $u_r \in \{1, \dots, U_r\}$ and $u_c \in \{1, \dots, U_c\}$.

\vspace{-6pt}
\subsection{MIS Equivalent Phase Shift Coefficients}
We designate MS1 as the primary layer and MS2 as the secondary layer of the MIS structure. Signals transmitted from the BS first arrive at MS1, sequentially pass through MS1 and MS2, and subsequently propagate to users. Given the close stacking and minimal thickness of both layers, signal leakage or additional phase deviation through non-overlapped elements is considered negligible.

Note that signals may refract either through the overlapped regions of both layers or through the non-overlapping part of MS1 alone. The model of MS1's static phase shifts mirrors a traditional RIS, while modeling the position-dependent effect of the movable MS2 is crucial. Therefore, to explicitly characterize the interaction
between static phase elements with dynamic geometry, we define an equivalent phase shift vector for the $u$-th beam pattern as:
\vspace{-3pt}
\begin{equation}
\boldsymbol{\bar{\theta}}_{u}= \boldsymbol{S}_u \boldsymbol{\theta} + \boldsymbol{e}_u \in \mathbb{C}^{M\times 1}, \forall u \in \mathcal{U}, 
\end{equation}
where $\boldsymbol{S}_u \in {0,1}^{M \times N}$ is a binary selection matrix, and $\boldsymbol{e}_u\in{0,1}^{M\times 1}$ is a binary padding vector defined as:
\vspace{-3pt}
\begin{subequations}
\begin{align}
\left[ \boldsymbol{S}_u \right] _{m,n}&=\begin{cases}
	1, &	\!\!\!\begin{array}{c}
	\text{if }n\text{-th element of MS2 locates}\\
	\!\!\!\!\!\!\!\text{upon }m\text{-th element of MS1},\\
\end{array}\\
	0,&		\text{otherwise},\\
\end{cases}
\\
\left[ \boldsymbol{e}_u \right]_m&=\begin{cases}
	0, &	\!\!\!\begin{array}{c}
	\text{if }n\text{-th element of MS2 locates}\\
	\!\!\!\!\!\!\!\text{upon }m\text{-th element of MS1},\\
\end{array}\\
	1,&		\text{otherwise}.\\
\end{cases}
\end{align}
\end{subequations}
Specifically, the selection matrix $\boldsymbol{S}_u$ maps MS2 elements onto corresponding positions of MS1, defining their overlapping pattern. The padding vector $\boldsymbol{e}_u$ assigns virtual MS2 elements with unit amplitude and zero phases to the MS1 elements that are not overlapped by MS2. This compact representation aligns the effective dimensions of both MS layers, unifying phase shift and transmission signal modeling across the two layers.

\vspace{-9pt}
\subsection{Spatially Correlated Rician Fading Channel Model}
The channels involved in the system include the BS-MIS channel matrix $\boldsymbol{G} \in \mathbb{C}^{M \times L}$, describing the complex coefficients between the $l$-th BS antenna and the $m$-th element of MS1, and the MIS-user channel vectors $\boldsymbol{h}_k \in \mathbb{C}^{M \times 1}, \forall k \in \mathcal{K}$, representing complex coefficients from the MIS elements (including MS1 and MS2) to the $k$-th user. Since BS and MIS are configured as UPAs with closely spaced antennas or elements, spatial correlation becomes significant, influencing achievable communication performance [29]. Accordingly, both the BS-MIS and MIS-user links are modeled as spatially correlated Rician fading channels
\begin{subequations}\label{eq:channel_model}
    \begin{align}
        \boldsymbol{h}_{k} &= \sqrt{\alpha_{2,k}} \bigg( \sqrt{\frac{\beta_{2,k}}{\beta_{2,k}+1}} \boldsymbol{\bar{h}}_{k} + \sqrt{\frac{1}{\beta_{2,k}+1}} \boldsymbol{\tilde{h}}_{k} \bigg), \\
        \boldsymbol{G} &= \sqrt{\alpha_1} \bigg( \sqrt{\frac{\beta_1}{\beta_1+1}} \boldsymbol{\bar{G}} + \sqrt{\frac{1}{\beta_1+1}} \boldsymbol{\tilde{G}} \bigg),  \end{align}
\end{subequations}
where $\alpha_{1}$ and $\alpha_{2,k}$ represent path loss coefficients for the BS-MIS and MIS-user $k$ links, respectively, while $\beta_{1}, \beta_{2,k}$ denote the corresponding Rician factors. The deterministic LoS components $\boldsymbol{\bar{G}}=\boldsymbol{a}_{\text{MIS}}(\vartheta_{\text{MIS}},\psi_{\text{MIS}})\boldsymbol{a}_{\text{BS}}(\vartheta_{\text{BS}},\psi_{\text{BS}})^T$ and $\boldsymbol{\bar{h}}_{k}=\boldsymbol{a}_k(\vartheta_k,\psi_k)$ are constructed from the array response vectors determined by azimuth and elevation angles $(\vartheta, \psi)$ at the BS, MIS, and users.

The NLoS components follow the correlated Rayleigh fading model via the Kronecker correlation structure\footnote{The fully passive MIS has pre-designed phases for quasi-static deployments leveraging statistical CSI. This assumption is consistent with the channel modeling with spatial consistency, where the statistics of correlation distances evolve smoothly over tens of meters rather than slot-by-slot [31]. The spatial correlated Rician model with a Kronecker NLoS structure captures finite angular spreads, departing from pure-LoS or oversimplified i.i.d. scattering. Obtaining $\mathbf{\hat{S}}_{\text{b}}$ follows the standard massive-MIMO procedures, while MIS-side user-specific spatial correlation are recoverable from the cascaded link via: (i) estimate dominant AoA/AoD and angular spreads from initial sounding, then synthesize array-consistent covariances; (ii) measure receive sample covariances at the BS and compact uesr feedback under a small set of representative MIS positions/patterns, and least-squares map them to $(\mathbf{\hat{S}}_{\text{mr}},\mathbf{\hat{S}}_{\text{mt}})$ pair; (iii) assisted initialization using ray-tracing or radio maps to reduce probing. Note that user-specific differences can be handled by a small set of user-cluster templates; our design can be extended to consume group-level statistical CSI, still avoiding per-slot, per-user covariance.} [30]
\vspace{-6pt}
\begin{subequations}
\begin{align}
        \boldsymbol{\tilde{h}}_{k} &= \boldsymbol{S}_{\text{mt}}^{\frac{1}{2}} \boldsymbol{z}_k, \forall k \in \mathcal{K},\\
        \boldsymbol{\tilde{G}} &= \boldsymbol{S}_{\text{mr}}^{\frac{1}{2}} \boldsymbol{\varSigma} \boldsymbol{S}_{\text{b}}^{\frac{1}{2}}, 
\end{align}
\end{subequations}    
where entries of $\boldsymbol{\varSigma}$ and $\boldsymbol{z}_k$ are independent identically distributed complex Gaussian variables, i.e., $\text{vec}\left( \boldsymbol{\varSigma } \right) \sim \mathcal{C}\mathcal{N}\left( \mathbf{0}_{L\times M},\boldsymbol{I}_L\otimes \boldsymbol{I}_M \right)$ and $\text{vec}\left( \boldsymbol{z}_k \right) \sim \mathcal{C}\mathcal{N}\left( \mathbf{0}_M,\boldsymbol{I}_M \right)$. The deterministic covariance matrices $\boldsymbol{S}_{\text{mt}}$, $\boldsymbol{S}_{\text{mr}}$, and $\boldsymbol{S}_{\text{b}}$ characterize spatial correlations between transmit antennas, receive antennas, and MIS elements, respectively, with each entry defined as:
\begin{subequations}
\begin{align}
&\left[ \boldsymbol{S}_{\text{mr}} \right] _{l,l^{'}}=\text{sinc}\Big( \frac{2\lVert \boldsymbol{u}_{\text{b},l}-\boldsymbol{u}_{\text{b},l^{'}} \rVert}{\lambda} \Big) , \forall l,l^{'}\in \mathcal{L},
\\
&\left[ \boldsymbol{S}_{\text{b}} \right] _{\text{m},m^{'}}=\text{sinc}\Big( \frac{2\lVert \boldsymbol{u}_{\text{mr},\text{m}}-\boldsymbol{u}_{\text{mr},m^{'}} \rVert}{\lambda} \Big) , \forall m,m^{'}\in \mathcal{M},
\\
&\left[ \boldsymbol{S}_{\text{mt}} \right] _{\text{m},m^{'}}=\text{sinc}\Big( \frac{2\lVert \boldsymbol{u}_{\text{mt},\text{m}}-\boldsymbol{u}_{\text{mt},m^{'}} \rVert}{\lambda} \Big) , \forall m,m^{'}\in \mathcal{M},
\end{align}
\end{subequations}
where $\lambda$ is the wavelength, and $\boldsymbol{u}_{\text{b},l}$, $\boldsymbol{u}_{\text{mr},m}$, and $\boldsymbol{u}_{\text{mt},m}$ represent the positions of the $l$-th BS antenna, $m$-th MS1 element, and $m$-th MS2 element, respectively. The notation $\|\cdot\|$ denotes the Euclidean distance between two elements.

\vspace{-6pt}
\subsection{Transmission Signal Model}
Building on the MIS architecture and channel model, we now present the transmission signal model. Recall that the MIS employs pre-designed static phase profiles, which should be optimized from long-term statistical CSI in the following. This makes it particularly suitable for quasi-static deployments such as industrial IoT, smart home, and sensor networks, where large-scale channel statistics remain stable and communication demands change periodically. Under these conditions, the overhead of fine-grained multiuser multiplexing offers little benefit.
Thus, we adopt a time-division multiple-access (TDMA) protocol to sequentially schedule different users. This TDMA approach aligns with the MIS operating principle of switching among a finite set of beam patterns via mechanical repositioning, while emphasizing our focus on the joint optimization of static phase shifts and beam pattern scheduling. Extensions to more sophisticated multiple-access schemes are readily conceivable and left for our future work.

When serving user $k$ with beam pattern $u$, the transmitted signal is $\boldsymbol{w}_{k,u}s_k$, where $\boldsymbol{w}_{k,u}\in\mathbb{C}^{M\times 1}$ is the BS beamforming vector, and $s_k\in\mathbb{C}$ is the data symbol with normalized power $\mathbb{E}[|s_k|^2]=1$. Accordingly, the received signal at user $k$ using beam pattern $u$ is expressed as\footnote{For general stacked MSs, we can write the two-layer effect as $\mathbf P=\boldsymbol\Phi_{2}\mathbf T_{2}\boldsymbol\Phi_{1}\mathbf T_{1}$ [32], where $\boldsymbol{T}_2$ is the transmission coefficient matrix between MS1 and MS2, while $\boldsymbol{T}_1$ is the matrix from the transmit antenna array to MS1. According to the MIS signal model, we have $\boldsymbol{\varPhi }_2=\mathrm{diag}\left( \boldsymbol{S}_{u}^{}\boldsymbol{\theta }^{*} \right) , \boldsymbol{\varPhi }_1=\mathrm{diag}\left( \boldsymbol{\phi }^{*} \right) , \boldsymbol{T}_1=\boldsymbol{G}^H,\boldsymbol{T}_2=c_2\left( g \right) \boldsymbol{I}$. With a tightly stacked gap $g\ll\lambda$ for MIS, which only leaves a minimal space to slide MS2, we model the inter-surface operator by a scaled identity $\mathbf T_{2}\approx c_{12}(g)\mathbf I$. Rationale: (i) for non-coaxial cell pairs, the Rayleigh–Sommerfeld (RS) kernel gives $t_{m,n} \propto \frac{g}{(\Delta\rho^{2}+g^{2})}\xrightarrow[g\to0]{}0$, so off-axis terms are negligible; (ii) for the coaxial pair the RS kernel is not physically valid at $g\to0$, thus we use an electromagnetic-consistent cascaded-sheet reduction wave-matrix: under normal incidence and polarization-independent sheets and spacers, the cascade acts as a scalar in each polarization, yielding $|c_{2}(g)|\leq 1$ [33]. In our baseband model, this can be absorb by the binary selection matrix $\boldsymbol{S}_u$ for signals through overlapped part of the MIS.}
\begin{subequations}
\begin{align}
y_k&=\boldsymbol{h}_{k}^{H}\mathrm{diag}( \boldsymbol{\bar{\theta}}_{u}^{H} ) \mathrm{diag}( \boldsymbol{\phi }^H ) \boldsymbol{G}\boldsymbol{w}_{k,u}s_k+n_k
\\
&=( \boldsymbol{\bar{\theta}}_{u}^{}\odot \boldsymbol{\phi } ) ^H\mathrm{diag}( \boldsymbol{h}_{k}^{H}) \boldsymbol{G}\boldsymbol{w}_{k,u}s_k+n_k, 
\end{align}
\end{subequations}
where $n_k \sim \mathcal{CN}(0, \sigma^2_k)$ is additive white Gaussian noise (AWGN) in the user $k$ and $\odot$ denotes the Hadamard product. 

\vspace{-6pt}
\section{Problem Formulation}
\vspace{-2pt}

This section formulates the joint optimization problem of static phase shift design and beam-pattern scheduling for MIS-assisted systems. We first describe a two-timescale design principle, distinguishing between statistical and instantaneous CSI. Then, we formulate optimization problems that aim to maximize the minimum user data rate and the total throughput.

\vspace{-7pt}
\subsection{Two-Timescale Design and Operation}
\vspace{-3pt}
Given an appropriately designed MIS, the BS adopts maximum-ratio transmission (MRT) beamforming in each time slot, leveraging instantaneous CSI to address the Rayleigh fading components and maximize the array gain. Specifically, the MRT beamforming vector for user $k$ when MIS beam pattern $u$ activates is defined as
\begin{align}
\boldsymbol{w}_{k,u}^{\text{MRT}}=\sqrt{P}\frac{\boldsymbol{G}^H\mathrm{diag}\left( \boldsymbol{h}_{k}^{} \right) \left( \boldsymbol{\bar{\theta}}_{u}^{}\odot \boldsymbol{\phi } \right)}{\lVert \boldsymbol{G}^H\mathrm{diag}\left( \boldsymbol{h}_{k}^{} \right) \left( \boldsymbol{\bar{\theta}}_{u}^{}\odot \boldsymbol{\phi } \right) \rVert}.
\end{align}
This setup naturally leads to a two-timescale design: instantaneous CSI is utilized at the short timescale for BS beamforming decisions, whereas the MIS static phase shifts and the beam pattern scheduling strategies rely on statistical CSI optimized over a longer timescale [14]. The resulting SNR at user $k$ served by beam pattern $u$ is thus expressed as
\begin{align}
\gamma_{k,u} =\iota_{k}\lVert \boldsymbol{G}^H\mathrm{diag}\left( \boldsymbol{h}_{k}^{} \right) \left( \boldsymbol{\bar{\theta}}_{u}^{}\odot \boldsymbol{\phi } \right) \rVert ^2,
\end{align}
where we denote $\iota_{k}=\frac{P}{\sigma _{k}^{2}}$ as the reference SNR of the $k$-th user taking into account transmit power $P$.

\vspace{-7pt}
\subsection{Minimum Data Rate Maximization}
\vspace{-3pt}
The MIS dynamically switches among the available beam patterns to serve different users. Due to its mechanical reconfiguration, only one beam pattern of MIS can be activated during each time slot. Our objective is to maximize the minimum data rate among all users, defined as
\begin{align}
\eta =\underset{k\in \mathcal{K}}{\min}\,\,\sum_{u\in \mathcal{U}}{\mathbb{E}\left\{ \log _2\left( 1+\xi_{k,u} \gamma _{k,u} \right) \right\}},
\end{align}
where $\xi_{k,u }\in \{0,1\}$ is the binary beam pattern scheduling variable with $\xi_{k,u} = 1$ indicating that $k$-th user is assigned to $u$-th beam pattern\footnote{MIS steers beams via mechanical position shifts between two fully passive static-phase surfaces, so it is not per-symbol reconfigured but targets quasi-static deployments where re-pointing is infrequent, e.g., seconds to minutes, with design based on long-term statistical CSI. Compared with electronically tunable RIS, MIS removes per-element phase shifters and dense cabling, cutting hardware complexity and continuous standby power consumption. While offering a finite beam codebook rather than frame-level agility, MIS achieves the intended cost–performance balance, well suited to coverage extension, periodic steering, and virtual-LoS establishment in quasi-static scenarios.}. With TDMA scheduling, each user $k$ is assigned exactly one optimal beam pattern $u$ that maximizes its ergodic data rate, while allowing multiple users to select a common beam pattern as needed. Consequently, the minimum-rate maximization problem (P1) can be formulated as
\begin{subequations}
\begin{align}
\text{(P1)}:&\underset{\eta ,\boldsymbol{\theta },\boldsymbol{\phi },\{\xi_{k,u}\}}{\max}\,\,\eta 
\\
\,\,\text{s}.\text{t}.\,\,&\sum_{u\in \mathcal{U}}{\xi_{k,u}\mathbb{E}\left\{ \log _2\left( 1+\gamma _{k,u} \right) \right\}} \ge \eta \,\,,\forall k\in \mathcal{K},
\\
&\xi _{k,u}\in \left\{ 0,1 \right\} ,\,\forall k\in \mathcal{K}, \forall u\in \mathcal{U},
\\
& \sum_{u\in \mathcal{U}}{\xi _{k,u}} = 1, \forall k\in \mathcal{K},
\\
&|\phi _m|=1,\,\forall m\in \mathcal{M},
\\
&|\theta _n|=1,\,\forall n\in \mathcal{N},
\end{align}
\end{subequations}
where constraint (10b) ensures each user’s minimum data rate, (10c) and (10d) enforce binary beam pattern assignments with exactly one beam pattern per user, and constraints (10e) and (10f) enforce the unit-modulus condition for MIS phase shifts.

\vspace{-6pt}
\subsection{Total Throughput Maximization}
\vspace{-2pt}
We further consider maximizing the total system throughput as problem (P2):
\begin{subequations}
\begin{align}
\text{(P2)} &:\underset{\eta ,\boldsymbol{\theta },\boldsymbol{\phi },\{\xi_{k,u}\}}{\max}\,\,\,\frac{T}{K}\sum_{k\in \mathcal{K}}{\sum_{u\in \mathcal{U}}{\xi_{k,u}\mathbb{E}\left\{ \log _2\left( 1+\gamma _{k,u} \right) \right\}}}
\\
\,\,\text{s.t.}\,\,& \text{(10c),(10d),(10e),(10f)}, \nonumber
\end{align}
\end{subequations}
where constraints are identical to those defined earlier. Both (P1) and (P2) represent mixed-integer non-convex optimization problems, which are typically NP-hard and non-trivial to solve optimally. To handle these complexities, we develop efficient solution approaches in Sections IV and V. Additionally, we generalize the MIS architecture by introducing element-wise mobility of MS2. Due to the structural difference between the resulting formulation and the block-level formulations (P1) and (P2), we separately address their formulation and solution methods in Section VI.

\section{Minimum Data Rate Maximization}
In this section, we address the minimum data rate maximization problem (P1). We first derive an analytical upper bound for each user's ergodic rate to obtain a tractable solution. Then, we reformulate the original mixed-integer non-convex optimization into a series of convex subproblems, which are solved iteratively using the BCD procedure.

\subsection{Ergodic Rate Upper Bound and Constraint Relaxation}

For user $k$ served by beam pattern $u$, Jensen's inequality yields the following upper bound on the ergodic rate\footnote{We replace $\mathbb{E}\{\log_2(1+\gamma)\}$ with $\log_2(1+\mathbb{E}\{\gamma]\})$ to obtain a tractable, deterministic metric under statistical CSI for quasi-/semi-static MIS. The Jensen gap is explicitly bounded as
$0\le \log_2(1+\mathbb{E}[Y])-\mathbb{E}[\log_2(1+Y)]\le \tfrac{\mathrm{Var}(Y)}{2\ln 2,(1+a)^2}$, where $a=\mathrm{ess,inf} \, Y\ge 0$ is the essential infimum of a random variable $Y$ and $\mathrm{Var}(\cdot)$ is the variance operator. In our setting $Y=\gamma_{k,u}$, and both $\mathbb{E}\{Y\}$ and $\mathbb{E}\{Y^2\}$ admit closed forms, so the bound is computable a priori. For Rician channels with moderate to high Rician factors and spatial correlation, $\mathrm{Var}(\gamma_{k,u})$ is small, yielding a tight bound; maximizing the surrogate consistently produces designs whose realized ergodic rates closely track it, as shown in the simulation section. We do not claim optimality for the original objective; rather, we provide a provable gap bound plus empirical evidence that the surrogate is decision-aligned in the targeted scenarios.} [34][35]:
\begin{align}
&\mathbb{E}\{ \log_2( 1+\gamma _{k,u}) \} \le \log_2( 1+\mathbb{E}\{ \gamma _{k,u} \}), 
\end{align}
where the expectation term for is calculated explicitly as
\begin{align}
\mathbb{E}\left\{ \gamma _{k,u} \right\} &=\iota_k \mathbb{E}\left\{ \lVert \boldsymbol{G}^H\mathrm{diag}\left( \boldsymbol{h}_k \right) \boldsymbol{v}_u \rVert ^2 \right\}  \nonumber
\\
&=\boldsymbol{v}_{u}^{H}\boldsymbol{\varXi }_k\boldsymbol{v}_u
\end{align}
where we denote $\boldsymbol{v}_u=\boldsymbol{\bar{\theta}}_{u}^{}\odot \boldsymbol{\phi }$, and $\boldsymbol{\varXi }_k$ is a positive semi-definite matrix given by
\begin{align}
\boldsymbol{\varXi}_k=&\iota_k\alpha_{1} \alpha_{2,k}\frac{\beta _{2,k}}{\beta _{2,k}+1}\frac{\beta _1}{\beta _1+1}\mathrm{diag} ( \boldsymbol{\bar{h}}^* ) \boldsymbol{\bar{G}}\boldsymbol{\bar{G}}^H\mathrm{diag}( \boldsymbol{\bar{h}} ) \nonumber
\\
+\iota_k&\alpha_{1} \alpha_{2,k}\frac{1}{\beta _{2,k}+1}\frac{\beta _1}{\beta _1+1}( \boldsymbol{S}_{\text{mt}}^{T}\odot ( \boldsymbol{\bar{G}}\boldsymbol{\bar{G}}^H) ) \nonumber
\\
+\iota_k&\alpha_{1} \alpha_{2,k}\frac{\beta _{2,k}}{\beta _{2,k}+1}\frac{1}{\beta _1+1}\text{tr}( \boldsymbol{S}_{\text{b}}^{} ) \cdot \mathrm{diag}( \boldsymbol{\bar{h}}^* ) \boldsymbol{S}_{\text{mr}}^{}\mathrm{diag} ( \boldsymbol{\bar{h}} ) \nonumber
\\
+\iota_k&\alpha_{1} \alpha_{2,k}\frac{1}{\beta _{2,k}+1}\frac{1}{\beta _1+1}\text{tr}\left( \boldsymbol{S}_{\text{b}}^{} \right) \cdot ( \boldsymbol{S}_{\text{mt}}^{T}\odot \boldsymbol{S}_{\text{mr}}^{}). 
\end{align}

Next, we relax the non-convex unit-modulus constraints $|\phi_m|=1$ and $|\theta_n|=1$ to convex constraints $|\phi_m|\leq 1$ and $|\theta_n|\leq 1$, respectively, and introduce a slack variable $\mu \geq 0$ to represent a common SNR lower bound. As such, an upper bound of the optimal value of (P1) can be obtained by solving the following problem
\begin{subequations}
\begin{align}
\text{(P3)} :&\underset{\boldsymbol{\theta },\boldsymbol{\phi },\{\xi_{k,u}\} ,\mu}{\max}\,\,\mu 
\\
\,\,\text{s.t.}\,\,&\sum_{u\in \mathcal{U}}{\xi _{k,u}^{}\boldsymbol{v}_{u}^{H}\boldsymbol{\varXi }_k\boldsymbol{v}_u}\ge \mu \,\,,\forall k\in \mathcal{K},
\\
&|\phi _m|\le 1,\,\forall m\in \mathcal{M},
\\
&|\theta _n|\le 1,\,\forall n\in \mathcal{N},
\\
&\text{(10c)},\text{(10d)}. \nonumber
\end{align}
\end{subequations}
The optimal solution $\mu^*$ from (P3) gives the upper-bound minimum ergodic rate as $\eta^*=\log_2(1+\mu^*)$.

\vspace{-12pt}
\subsection{Penalty-Based Problem Reformulation}
The binary constraints $\xi_{k,u}\in\{0,1\}$ in (10c) hinder direct solutions to (P3). To handle this, we rewrite the binary constraints equivalently as
\begin{subequations}
\begin{align}
&0\le \xi _{k,u}\le 1,\, \forall k\in \mathcal{K},\forall u\in \mathcal{U},
\\
&\xi _{k,u}-\xi _{k,u}^{2}\le 0,\, \forall k\in \mathcal{K},\forall u\in \mathcal{U},
\end{align}
\end{subequations}
where (16b) is a reverse-convex constraint leading to a disconnected feasible set. To circumvent this difficulty, we incorporate (16b) into the objective function as a penalty term, obtaining the reformulated problem
\begin{align}
\left( P4 \right) :&\underset{\boldsymbol{\theta },\boldsymbol{\phi },\{\xi_{k,u}\} ,\mu}{\max}\,\,\mu -\rho h\left( \left\{ \xi _{k,u} \right\} \right) 
\\
\,\,\text{s.t.}\,\,& \text{(10d), (15b), (15c), (15d), (16a)},  \nonumber
\end{align}
where $h(\{ \xi _{k,u} \} ) \triangleq \sum_{u\in \mathcal{U}}{\sum_{k\in \mathcal{K}}{(\xi _{k,u}-\xi _{k,u}^{2})}}$ and $\rho \ge 0$ serves as a penalty parameter to penalize the violation of constraint (16b). To maximize the objective function of (P4) when $\rho \rightarrow \infty$, the optimal $\{ \xi _{k,u}^{*}\}$ should meet the condition
$h(\{ \xi _{k,u}^{*}\}) \leq 0$. On the other hand, since $\{ \xi _{k,u}^{*}\}$ satisfies the constraint (16a), we have $h(\{ \xi _{k,u}^{*}\}) \geq 0$. Thus, $h(\{ \xi _{k,u}^{*}\})=0$ and accordingly $\xi _{k,u}^{*}\in \{0,1\}, \forall k\in \mathcal{K}, \forall u \in \mathcal{U}$ follows, which verifies the equivalence between problems (P3) and (P4). It is worth mentioning that since setting $\rho$ significantly large at the beginning can render this approach ineffective, we initialize $\rho$ to a small value to find a good starting point and then solve problem (P4) iteratively with $\rho$ increasing with iterations until $h(\{\xi_{k,u}^*\})\rightarrow 0$ [36].

For any given $\rho$, problem (P4) remains non-convex and still hard to solve directly due to the non-concave objective function and the non-convex constraints with closely coupled variables in (15b). Nevertheless, observing that the problem becomes more tractable when optimizing one variable block at a time, we resort to the BCD method, iteratively updating the beam pattern scheduling variables $\{\xi_{k,u}\}$ and the MIS static phase shifts $\boldsymbol{\phi}$ and $\boldsymbol{\theta}$, as detailed below.

\vspace{-12pt}
\subsection{Optimizing $\{\xi_{k,u}\}$ for Given $\boldsymbol{\phi}$ and $\boldsymbol{\theta}$}

Given fixed MIS phase shifts, the scheduling variables $\{\xi_{k,u}\}$ are optimized by solving the following subproblem
\begin{align}
\left( P4.1 \right) :&\underset{\mu,\{\xi_{k,u}\}}{\max}\,\,\mu -\rho h\left( \left\{ \xi _{k,u} \right\} \right) 
\\
\,\,\text{s.t.}\,\,& \text{(10d), (15b), (16a)},  \nonumber
\end{align}
where the constraints are linear. The convex penalty term $h\left( \left\{ \xi _{k,u} \right\} \right)$ makes the objective function non-concave, leading to the non-convexity of (P4.1).
To handle this problem, we leverage the iterative SCA technique. 

Specifically, since the first-order Taylor expansion of any convex (concave) function at any point is its global lower (upper) bound, the following inequalities hold:
\begin{align}
\xi _{k,u}^{2}\ge -\left( \xi _{k,u}^{\ell} \right) ^2+2\xi _{k,u}^{\ell}\xi _{k,u}^{}\triangleq \varsigma ^{\text{lb},\ell}\left( \xi _{k,u}^{} \right),
\end{align}
where $\xi_{k,u}^\ell$ is the given local point in the $\ell$-th iteration of SCA. In this way, we can replace the penalty function with its first-order Taylor expansion-based linear upper bound
\begin{align}
h^{\text{ub},\ell}\left( \left\{ \xi _{k,u}^{} \right\} \right) =\sum_{k\in \mathcal{K}}{\sum_{u\in \mathcal{U}}{\left( \xi _{k,u}^{}-\varsigma ^{\text{lb},\ell}\left( \xi _{k,u}^{} \right) \right)}},
\end{align}
Then, we convert problem (P4.1) to a convex linear program
\begin{align}
\left( P4.2 \right) :&\underset{\mu,\{\xi_{k,u}\}}{\max}\,\,\mu -\rho h^{\text{ub},\ell}\left( \left\{ \xi _{k,u}^{} \right\} \right) 
\\
\,\,\text{s.t.}\,\,& \text{(10d), (15b), (16a)},  \nonumber
\end{align}
and the optimal solution (P4.2) can be obtained by resorting to standard convex solvers such as CVX [37].

\vspace{-12pt}
\subsection{Optimizing $\boldsymbol{\phi}$ for Given $\{\xi_{k,u}\}$ and $\boldsymbol{\theta}$}

Given fixed scheduling decisions $\{\xi_{k,u}\}$, optimization of $\boldsymbol{\phi}$ or $\boldsymbol{\theta}$ separately results in non-convex quadratic constraints. Before focusing on the optimization algorithm, we rewrite terms in (13) related to the MIS phase shifts in explicit forms with respect to (w.r.t.) $\boldsymbol{\theta}$ and $\boldsymbol{\phi}$, respectively, as below.
Combined (1) with (13), we have
\begin{subequations}
\begin{align}
&\left( \boldsymbol{\bar{\theta}}_u\odot \boldsymbol{\phi } \right) ^H\boldsymbol{\varXi }_k\left( \boldsymbol{\bar{\theta}}_u\odot \boldsymbol{\phi } \right)  \nonumber
\\
&=\left( \boldsymbol{S}_u\boldsymbol{\theta }\odot \boldsymbol{\phi } \right)^H \! \boldsymbol{\varXi }_k \! \left( \boldsymbol{S}_u\boldsymbol{\theta }\odot \boldsymbol{\phi } \right) + \left( \boldsymbol{S}_u\boldsymbol{\theta }\odot \boldsymbol{\phi } \right) ^H \! \boldsymbol{\varXi }_k \! \left( \boldsymbol{e}_u\odot \boldsymbol{\phi } \right)  \nonumber
\\
&\quad+\left( \boldsymbol{e}_u\odot \boldsymbol{\phi } \right) ^H\boldsymbol{\varXi }_k\left( \boldsymbol{S}_u\boldsymbol{\theta }\odot \boldsymbol{\phi } \right) +\left( \boldsymbol{e}_u\odot \boldsymbol{\phi } \right) ^H\boldsymbol{\varXi }_k\left( \boldsymbol{e}_u\odot \boldsymbol{\phi } \right)  \nonumber
\\
&=\boldsymbol{\theta }^H\boldsymbol{A}_{k,u}\boldsymbol{\theta }^H+2\text{Re}\{ \boldsymbol{\theta }^H\boldsymbol{a}_{k,u}\} +a_{k,u}
\\
&=\boldsymbol{\phi }^H\boldsymbol{B}_{k,u}\boldsymbol{\phi},
\end{align}
\end{subequations}
where (22a) and (22b) can be equivalently deemed as quadratic functions w.r.t. $\boldsymbol{\theta}$ and $\boldsymbol{\phi}$, respectively, with
\begin{subequations}
\begin{align}
&\boldsymbol{A}_{k,u}=\boldsymbol{S}_{u}^{H}\mathrm{diag}( \boldsymbol{\phi }^H ) \boldsymbol{\varXi }_k\mathrm{diag}( \boldsymbol{\phi } ) \boldsymbol{S}_u, 
\\
&\boldsymbol{a}_{k,u}=\boldsymbol{S}_{u}^{H}\mathrm{diag}( \boldsymbol{\phi }^H) \boldsymbol{\varXi }_k\mathrm{diag}( \boldsymbol{\phi } ) \boldsymbol{e}_u, 
\\
&a_{k,u}=\boldsymbol{e}_{u}^{H}\mathrm{diag}( \boldsymbol{\phi }^H ) \boldsymbol{\varXi }_k\mathrm{diag}( \boldsymbol{\phi } ) \boldsymbol{e}_u,
\\
&\boldsymbol{B}_{k,u}=\mathrm{diag}( \boldsymbol{\theta }^H\boldsymbol{S}_{u}^{H}+\boldsymbol{e}_{u}^{H} ) \boldsymbol{\varXi }_k\mathrm{diag}( \boldsymbol{S}_u\boldsymbol{\theta }+\boldsymbol{e}_u). 
\end{align}
\end{subequations}

Now, given any feasible $\{\xi_{k,u}\}$ and $\boldsymbol{\theta}$, by ignoring the constant term in the objective function of problem (P4), we can equivalently express the subproblem w.r.t. {$\boldsymbol{\phi}$} as
\begin{subequations}
\begin{align}
\text{(P4.3)}:&\underset{\mu,\boldsymbol{\phi }}{\max}\,\,\mu 
\\
\,\,\text{s.t.}\,\,&\sum_{u\in \mathcal{U}}{\xi _{k,u}^{}\boldsymbol{\phi }^H\boldsymbol{B}_{k,u}\boldsymbol{\phi }}\ge \mu \,\,,\forall k\in \mathcal{K},
\\
& \text{(15c)}. \nonumber
\end{align}
\end{subequations}
It is obvious that constraints (24b) are non-convex since the quadratic terms on the left-hand side of them are convex w.r.t. $\boldsymbol{\phi}$, which motivates us to convexify these two constraints using the SCA technique. To be specific, by replacing the left-hand side of (24b) with its respective first-order Taylor expansions at the given local points $\boldsymbol{\phi}^\ell$ in the $\ell$-th iteration of SCA, (24b) can be approximated as the following convex constraints:
\begin{align}
\sum_{u\in \mathcal{U}}{\xi_{k,u}\mathcal{F}_{\boldsymbol{B}_{k,u}}^{\text{lb},\ell}\left( \boldsymbol{\phi } \right)} \ge \mu _{k,u},\,\forall k\in \mathcal{K},
\end{align}
Then, a locally optimal solution of (P4.3) can be obtained by iteratively solving the following convex problem (P4.4) via readily available solvers (e.g., CVX [37])
\begin{align}
\text{(P4.4)}:\underset{\mu,\boldsymbol{\phi }}{\max}\,\,\mu  \qquad\text{s.t.}\,\,\text{(15c), (25)}. \nonumber
\end{align}

\vspace{-12pt}
\subsection{Optimizing $\boldsymbol{\theta}$ for Given $\{\xi_{k,u}\}$ and $\boldsymbol{\phi}$}

Similarly to the process in the subproblem (P4.3) and (P4.4), given any feasible $\{\xi_{k,u}\}$ and $\boldsymbol{\theta}$, by ignoring the constant term in the objective function of the problem (P4) and resorting to SCA to handle the quadratic term in (22a), i.e., $\boldsymbol{\theta}^H\boldsymbol{A}_{k,u}\boldsymbol{\theta }^H+2\text{Re}\{ \boldsymbol{\theta }^H\boldsymbol{a}_{k,u}\} +a_{k,u}$, we can express the convexified subproblem w.r.t. $\boldsymbol{\theta}$  as
\begin{subequations}
\begin{align}
\text{(P4.5)}:&\underset{\mu,\boldsymbol{\theta}}{\max}\,\,\mu 
\\
\,\,\text{s.t.}\,\,&\sum_{u\in \mathcal{U}}{\xi_{k,u}\mathcal{F}_{\boldsymbol{A}_{k,u}}^{\text{lb},\ell}\left( \boldsymbol{\theta } \right)}\ge \mu \,\,,\forall k\in \mathcal{K},
\\
& \text{(15d)}, \nonumber
\end{align}
\end{subequations}
where $\mathcal{F}_{\boldsymbol{A}}^{\text{lb}}$ is the first-order Taylor expansion-based lower bound of (24b)
\begin{align}
\mathcal{F}_{\boldsymbol{A}}^{\text{lb}}( \boldsymbol{\theta}^{\ell}) =&2\text{Re}\{ \boldsymbol{\theta}^H \boldsymbol{A}_{k,u} \boldsymbol{\theta}^{\ell} \} -( \boldsymbol{\theta}^{\ell} ) ^H\boldsymbol{A}_{k,u}\boldsymbol{\theta}^{\ell} \nonumber
\\
&+2\text{Re}\{ \boldsymbol{\theta }^H\boldsymbol{a}_{k,u} \} +a_{k,u}.
\end{align}

\vspace{-12pt}
\subsection{Overall Algorithm}
Algorithm 1 summarizes the overall iterative BCD procedure for (P1). For any given $\rho$, the BCD inner loop of Algorithm 1 solves problem (P4) by alternately solving problems (P4.2), (P4.4), and (P4.5), which is guaranteed to converge to a stationary point of the problem (P4) [38]. In the outer loop, we gradually increase $\rho$ to a sufficiently large value through $\rho \leftarrow \zeta \rho$ to make $h\big(\{a_{k,\ell}\}\big) \rightarrow 0$, thus ensuring $\xi_{k,u}^\ell \in \{0,1\}$, $\forall k\in\mathcal K, \ell\in\mathcal L$. As a consequence, after the convergence of the outer loop, we can obtain a stationary solution of problem (P3) satisfying the binary constraints on $\{a_{k,\ell}\}$. Then, by performing the remaining operations in Steps 12 and 13 of Algorithm 1, we can obtain a suboptimal solution of (P1). 

The dominant computational cost occurs in steps 5–7, where the optimization subproblems are solved iteratively. Step 5 involves solving a linear program (P4.2) with $KU$ scheduling variables. Using an interior-point solver, the complexity is approximately $\mathcal{O}(K^3 U^3)$. Step 6 involves optimizing the static phase shift vector with $M$ elements, formulated as a convex quadratic problem, with complexity via an interior-point solver being approximately $\mathcal{O}\left( M ^3 \right)$. Similarly, Step 7 yields complexity of $\mathcal{O}\left(N^3 \right)$. Considering the number of inner-loop iterations $I_{\rm inn}$ required for convergence to a stationary point and the outer-loop penalty iterations $I_{\rm out}$ for enforcing binary constraints, the overall computational complexity of Algorithm 1 is $\mathcal{O}( I_{\text{in}}I_{\text{out}}( \left( KU \right) ^3+M^3+N^3 ) ) $\footnote{The optimization algorithm for MIS is executed offline to co-design the static phase shifts and a finite beam-pattern codebook. After deployment, operation is two-timescale: fast-time communication uses standard BS precoding with instantaneous CSI, while slow-time reconfiguration selects a pattern index from the precomputed beam patterns and triggers a single mechanical shift of MS2. The run-time signaling overhead is only $\log_2|\mathcal{U}|$ bits per switch (searching over beam pattern set with size $|\mathcal{U}|$), and moves are infrequent driven by quasi-static changes such as hotspot relocation. Hence, the intensive computational complexity of optimization algorithms is amortized and decoupled from real-time operation. For applications requiring sub-second, per-frame agility, an electronically tunable RIS is the appropriate choice; for quasi-static deployments, MIS achieves the desired coverage steering with far lower hardware and continuous power cost.}.

\begin{algorithm}[t]
\caption{Proposed Algorithm for Problem (P1)}
\label{alg:optimization_algorithm}
\begin{algorithmic}[1]
    \STATE Initialize $\{\xi_{k,u}^0\}$, $\boldsymbol{\phi}^0$, $\boldsymbol{\theta}^0$, $\rho \ge 0$, and $\zeta\ge 1$. 
    \REPEAT 
        \STATE Set $\ell$=0, $\rho = \zeta \rho$
        \REPEAT
            \STATE Solve (P4.2) for given $\{\boldsymbol{\phi}^{\ell}\}$ and $\boldsymbol{\theta}^{\ell}$, and denote the obtained locally optimal solution as $\{\xi_{k,u}^{\ell+1}\}$.  
            \STATE Solve (P4.4) for given $\{\xi_{k,u}^{\ell}\}$ and $\boldsymbol{\theta}^{\ell}$, and denote the obtained locally optimal solution as $\boldsymbol{\phi}^{\ell+1}$.
            \STATE Solve (P4.5) for given $\{\xi_{k,u}^{\ell}\}$ and $\boldsymbol{\phi}^{\ell}$, and denote the obtained locally optimal solution as $\boldsymbol{\theta}^{\ell+1}$. 
            \STATE $\ell \gets \ell +1$.
        \UNTIL The fractional increase of the objective value of problem (P3) is smaller than a threshold $\epsilon_1 \ge 0$.
        \STATE $\{\xi_{k,u}^0\} \gets \{\xi_{k,u}^{\ell}\}$, $\boldsymbol{\phi}^0 \gets \boldsymbol{\phi}^{\ell}$, and $\boldsymbol{\theta}^0 \gets \boldsymbol{\theta}^{\ell}$
    \UNTIL $h\left( \{ \xi _{k,u}^\ell \} \right)$ is below a threshold $\epsilon_2 \ge 0$.
    \STATE Apply binary thresholding for $\{\xi_{k,u}\}$ to derive $\{\xi_{k,u}^*\}$ and set $[\boldsymbol{\theta}^*]_n=[\boldsymbol{\theta}^\ell]_n / |[\boldsymbol{\theta}^\ell]_n|$ and $[\boldsymbol{\phi}^*]_n=[\boldsymbol{\phi}^\ell]_n / |[\boldsymbol{\phi}^\ell]_n|$.
    \STATE Compute $\eta$ based on the optimized variables, and output $\{\eta^*,\{\xi_{k,u}^*\}, \boldsymbol{\phi}^*, \boldsymbol{\theta}^*\}$ as a suboptimal solution of (P1).
\end{algorithmic}
\end{algorithm}

\vspace{-6pt}
\section{Total Throughput Maximization}
\vspace{-3pt}
In this section, we focus on the total throughput maximization problem (P2). Although this problem could be addressed via a similar BCD-SCA method as described in the previous section, we instead employ manifold optimization techniques, which efficiently exploit the smoothness of the objective function and the inherent structure of the product manifold formed by the optimization variables.

\vspace{-6pt}
\subsection{Product Manifold Constructing}
\vspace{-3pt}
The phase shift vectors $\boldsymbol{\phi}$ and $\boldsymbol{\theta}$ are subject to unit-modulus constraints, i.e., $|\phi_m| = 1, \forall m \in \mathcal{M}$ and $|\theta_n| = 1, \forall n \in \mathcal{N}$, inherently defining a Riemannian manifold known as the \textit{complex circle manifold}. However, the beam pattern scheduling variables $\xi_{k,u}$ are originally binary and do not directly constitute a manifold. To tackle this difficulty, we first relax the discrete binary variables $\{\xi_{k,u}\}$ into continuous variables within the unit interval
\begin{align}
\xi_{k,u}\in \left\{ 0,1 \right\} \Rightarrow 1\ge \xi_{k,u}\ge 0, \,  \forall k \in \mathcal{K}, \forall u \in \mathcal{U}.
\end{align}
Taking into account the additional constraint $\sum_{u=1}^U{\xi_{k,u}}=1,\forall k \in \mathcal{K}$ in (10d) and enforcing strict positivity, the relaxed variables ${\xi_{k,u}}$ reside within a probability simplex, thus forming a \emph{multinomial manifold} [39]
\begin{align}
\sum_{u=1}^U{\xi_{k,u}}=1, \xi_{k,u} > 0, \,  \forall k \in \mathcal{K}, \forall u \in \mathcal{U}.
\end{align}
The structure of the objective function inherently drives these relaxed variables toward binary values at convergence. In particular, for each user $k$, the optimal $\xi_{k,u}$ becomes one for the beam pattern $u$ that achieves the highest rate, and zero otherwise. Therefore, the relaxation is tight in practice, as $\xi_{k,u}$ approaches binary values in the solution.

Consequently, we define the feasible sets of the optimization variables as the following manifolds
\begin{subequations}
\begin{align}
    &\mathcal{R}_{\boldsymbol{\phi}} = \left\{ \boldsymbol{\phi} \in \mathbb{C}^M : |\phi_m| = 1, \forall m \in \mathcal{M} \right\}, 
    \\
    &\mathcal{R}_{\boldsymbol{\theta}} = \left\{ \boldsymbol{\theta} \in \mathbb{C}^N : |\theta_n| = 1, \forall n \in \mathcal{N}\right\}, 
    \\
    &\mathcal{R}_{\boldsymbol{X}} \! =\! \Big\{ \!\boldsymbol{X} \in \mathbb{R}^{K \times U}\!:\! \sum_{u\in \mathcal{U}} \xi_{k,u}\! =\!1, \xi_{k,u} \!>\!0, \forall u \in \mathcal{U}, \forall k \in \mathcal{K}  \Big\}.
\end{align}
\end{subequations}
A manifold can be interpreted as a topological space that locally resembles a Euclidean space, and a tangent vector describes the direction in which a point can be updated on the manifold. All tangent vectors at a given point, representing all possible directions in which the point can move, collectively form the tangent space, which is here explicitly defined as:
\begin{subequations}
\begin{align}
&\mathcal{T}_{\boldsymbol{\phi }}=\big\{\boldsymbol{t}\in \mathbb{C}^M: \Re\left\{ \phi _m^* t_m \right\} =0,\forall m\in \mathcal{M}\big\},
\\
&\mathcal{T}_{\boldsymbol{\theta }}=\big\{\boldsymbol{t}\in \mathbb{C}^N: \Re\left\{ \theta _n^* t_n \right\} =0,\forall n\in \mathcal{N}\big\},
\end{align}
\begin{align}
&\mathcal{T}_{\boldsymbol{X}}=\Big\{ \boldsymbol{T}\in \mathbb{R}^{K\times U}: \sum_{u \in \mathcal{U}}{t_{k,u}}=0,\forall u\in \mathcal{U}, \forall k\in \mathcal{K} \Big\},\!\!
\end{align}
\end{subequations}
where condition $\sum_{u \in \mathcal{U}}{t_{k,u}}=0$ ensures that any infinitesimal perturbation within the tangent space does not violate the simplex constraint that the components sum to one.
Subsequently, by taking the Cartesian product of the individual manifolds for optimization variables, we construct a product manifold defined as 
\begin{align}
    \mathcal{R}_{(\boldsymbol{\phi},\boldsymbol{\theta},\boldsymbol{X})} = \mathcal{R}_{\boldsymbol{\phi}} \times \mathcal{R}_{\boldsymbol{\theta}} \times \mathcal{R}_{\boldsymbol{X}},
\end{align}
where $\times$ denotes the Cartesian product between sets. For any point $(\boldsymbol{\phi}, \boldsymbol{\theta}, \mathbf{X}) \in \mathcal{R}_{(\boldsymbol{\phi},\boldsymbol{\theta},\boldsymbol{X})}$, the tangent space at that point is the direct sum of the individual tangent spaces
\begin{align}
\mathcal{T}_{(\boldsymbol{\phi},\boldsymbol{\theta},\mathbf{X})} = \mathcal{T}_{\boldsymbol{\phi}} \oplus \mathcal{T}_{\boldsymbol{\theta}} \oplus \mathcal{T}_{\mathbf{X}},
\end{align}
where $\oplus$ denotes the direct sum of vector spaces.

\subsection{RCG Method}
With the product manifold clearly defined in (33), we reformulate (P2) as the following unconstrained Riemannian optimization problem
\begin{align}
\text{(P5):}\,\,\max_{\boldsymbol{\phi},\boldsymbol{\theta},\boldsymbol{X}\in\mathcal{R}_{(\boldsymbol{\phi},\boldsymbol{\theta},\boldsymbol{X})}}&\,\,f\left( \boldsymbol{\phi },\boldsymbol{\theta},\boldsymbol{X} \right), \nonumber
\end{align}
where
\begin{align}
f\left( \boldsymbol{\phi },\boldsymbol{\theta },\boldsymbol{X} \right) =\sum_{k\in \mathcal{K}}{\sum_{u\in \mathcal{U}}{\xi _{k,u}\log _2\left( 1+\iota _k\boldsymbol{v}_{u}^{H}\boldsymbol{\varXi }_k\boldsymbol{v}_u \right)}},
\end{align}
is the upper bound of the original objective function (11a) according to Jensen's equality used in (12) and (13) after ignoring the constant ratio $\frac{T}{K}$.

This formulation enables the joint and simultaneous optimization of all variables within a unified manifold framework, thereby circumventing the need for unit-modulus relaxation and the complexity of BCD procedures. As a result, the approach achieves improved computational efficiency by accounting for the interactions among all variables. To solve (P5), we adopt the Riemannian Conjugate Gradient (RCG) method, which generalizes the classical conjugate gradient algorithm to optimization over Riemannian manifolds (curved spaces). The RCG method proceeds by computing the Riemannian gradient, determining the conjugate search direction, and applying a retraction operation to update variables and ensure that updated variables remain on the manifold.

\subsubsection{Computing the Riemannian gradient}
Before determining the search direction, we need to calculate the Riemannian gradient, a vector field on the manifold $ \mathcal{R}$ obtained by projecting the Euclidean gradient onto the tangent space of the manifold. The explicit Euclidean gradients of $f(\boldsymbol{\phi}, \boldsymbol{\theta}, \boldsymbol{X})$ w.r.t. variables $\boldsymbol{\phi}$, $\boldsymbol{\theta}$, and $\boldsymbol{X}$ are given respectively by (36).
\begin{subequations}
\begin{figure*}
\begin{align}
&\qquad \qquad \nabla _{\boldsymbol{\phi }}f\left( \boldsymbol{\phi },\boldsymbol{\theta },\boldsymbol{X} \right) =\frac{2}{\ln 2}\sum_{k=1}^K{\sum_{u=1}^U{\frac{\xi _{k,u}}{1+\gamma _{k,u}\left( \boldsymbol{\phi },\boldsymbol{\theta } \right)}\mathrm{diag}\left( \boldsymbol{S}_u\boldsymbol{\theta }^*+\boldsymbol{e}_u \right) \boldsymbol{\varXi }_k\mathrm{diag}\left( \boldsymbol{S}_u\boldsymbol{\theta }+\boldsymbol{e}_u \right) \boldsymbol{\phi }}},
\\
&\qquad \qquad \nabla _{\boldsymbol{\theta }}f\left( \boldsymbol{\phi },\boldsymbol{\theta },\boldsymbol{X} \right) =\frac{2}{\ln 2}\sum_{k=1}^K{\sum_{u=1}^U{\frac{\xi _{k,u}}{1+\gamma _{k,u}\left( \boldsymbol{\phi },\boldsymbol{\theta } \right)}\boldsymbol{S}_{u}^{T}\mathrm{diag}\left( \boldsymbol{\phi }^* \right)}}\boldsymbol{\varXi }_k\mathrm{diag}\left( \boldsymbol{\phi } \right) \left( \boldsymbol{S}_u\boldsymbol{\theta }+\boldsymbol{e}_u \right),
\\
&\qquad \qquad \nabla _{\boldsymbol{X}}f\left( \boldsymbol{\phi },\boldsymbol{\theta },\boldsymbol{X} \right) =\left[ \log _2\left( 1+\gamma _{k,u}\left( \boldsymbol{\phi },\boldsymbol{\theta } \right) \right) \right] _{K\times U}.
\\
& \overline{\ \ \ \ \ \ \ \ \ \ \ \ \ \ \ \ \ \ \ \ \ \ \ \ \ \ \ \ \ \ \ \ \ \ \ \ \ \ \ \ \ \ \ \ \ \ \ \ \ \ \ \ \ \ \ \ \ \ \ \ \ \ \ \ \ \ \ \ \ \ \ \ \ \ \ \ \ \ \ \ \ \ \ \ \ \ \ \ \ \ \ \ \ \ \ \ \ \ \ \ \ \ \ \ \ \ \ \ \ \ \ \ \ \ \ \ \ \ \ \ \ \ \ \ \ \ \ \ \ \ \ \ \ \ \ \ \ \ \ \ \ \ \ \ \ } \nonumber
\end{align}
\vspace{-36pt}
\end{figure*}
\end{subequations}
Having computed the Euclidean gradients, we project them onto the tangent spaces of their respective manifolds to obtain the Riemannian gradients. Specifically, the Riemannian gradient $\nabla_{\mathcal{R}_{\boldsymbol{\phi}}} f(\boldsymbol{\phi}, \boldsymbol{\theta}, \boldsymbol{X})$ is obtained by projecting the Euclidean gradient $\nabla_{\boldsymbol{\phi}} f(\boldsymbol{\phi}, \boldsymbol{\theta}, \boldsymbol{X})$ onto the tangent space $\mathcal{T}_{\boldsymbol{\phi}}$
\begin{subequations}
\begin{align}
&\nabla_{\mathcal{R}_{\boldsymbol{\phi}} } f(\boldsymbol{\phi}, \boldsymbol{\theta}, \boldsymbol{X}) \nonumber
\\
&= \mathsf{Proj}_{\boldsymbol{\phi}}(\nabla_{\boldsymbol{\phi}} f(\boldsymbol{\phi}, \boldsymbol{\theta}, \boldsymbol{X})) 
\\
&= \nabla_{\boldsymbol{\phi}} f(\boldsymbol{\phi}, \boldsymbol{\theta}, \boldsymbol{X}) - \Re\left( \nabla_{\boldsymbol{\phi}} f(\boldsymbol{\phi}, \boldsymbol{\theta}, \boldsymbol{X}) \odot \boldsymbol{\phi}^* \right) \odot \boldsymbol{\phi},
\end{align}
\end{subequations}
where $\mathsf{Proj}_{\boldsymbol{\phi}}(\cdot)$ denotes the projection operation, $\odot$ denotes the Hadamard (element-wise) product, and $\boldsymbol{\phi}^*$ is the complex conjugate of $\boldsymbol{\phi}$.
Similarly, the Riemannian gradient $\nabla_{\mathcal{R}_{\boldsymbol{\theta}}} f(\boldsymbol{\phi}, \boldsymbol{\theta}, \boldsymbol{X})$ is given by
\begin{align}
\nabla_{\mathcal{R}_{\boldsymbol{\theta}} }& f(\boldsymbol{\phi}, \boldsymbol{\theta}, \boldsymbol{X}) \nonumber
\\
=& \nabla_{\boldsymbol{\theta}} f(\boldsymbol{\phi}, \boldsymbol{\theta}, \boldsymbol{X}) - \Re\left( \nabla_{\boldsymbol{\theta}} f(\boldsymbol{\phi}, \boldsymbol{\theta}, \boldsymbol{X}) \odot \boldsymbol{\theta}^* \right) \odot \boldsymbol{\theta}.
\end{align}
For the beam pattern scheduling matrix $\boldsymbol{X} \in \mathbb{R}^{K \times U}$, which lies on the multinomial manifold, the Riemannian gradient $\nabla_{\mathcal{R}_{\boldsymbol{X}}} f(\boldsymbol{\phi}, \boldsymbol{\theta}, \boldsymbol{X})$ is obtained by 
\begin{subequations}
\begin{align}
&\nabla_{\mathcal{R}_{\boldsymbol{X}} } f(\boldsymbol{\phi}, \boldsymbol{\theta}, \boldsymbol{X}) \nonumber
\\
&= \mathsf{Proj}_{\boldsymbol{X}}(\nabla_{\boldsymbol{X}} f(\boldsymbol{\phi}, \boldsymbol{\theta}, \boldsymbol{X})) 
\\
&= \nabla_{\boldsymbol{X}} f(\boldsymbol{\phi}, \boldsymbol{\theta}, \boldsymbol{X}) - \Big( \frac{1}{U}\nabla_{\boldsymbol{X}} f(\boldsymbol{\phi}, \boldsymbol{\theta}, \boldsymbol{X}) \boldsymbol{1}_{U\times1} \Big) \boldsymbol{1}_{U\times1}^\mathrm{T},\!\!
\end{align}
\end{subequations}
where $\boldsymbol{1}_{U\times1}$ is a column vector of length $U$, and the subtraction ensures that each row of $\nabla_{\mathcal{R}_{\boldsymbol{X}} } f(\boldsymbol{\phi}, \boldsymbol{\theta}, \boldsymbol{X})$ sums to zero, satisfying the tangent space conditions in (31c) of the multinomial manifold in a probability simplex form.

\subsubsection{Determining Conjugate Descent Direction}
To determine the descent direction on the manifold, we employ the conjugate gradient method adapted to Riemannian manifolds. Specifically, in iteration $i$, the descent direction $ \boldsymbol{\eta }^{\left( i \right)}$, $\boldsymbol{\tau }^{\left( i \right)}$, and $\boldsymbol{\varXi }^{\left( i \right)}$ for the variables $\boldsymbol{\phi}^{(i)}$, $\boldsymbol{\theta}^{(i)}$, and $\boldsymbol{X}^{(i)}$, respectively, are calculated using the Polak-Ribiere formula. For brevity, the function $f(\boldsymbol{\phi}^{(i)},\boldsymbol{\theta}^{(i)},\boldsymbol{X}^{(i)})$ is abbreviated as $f^{(i)}$ in the following.
Let $\nabla _{\mathcal{R}_{\boldsymbol{\phi }}}f^{(i)}$ denote the Riemannian gradient in iteration $i$. The conjugate gradient update coefficient $\beta_{\boldsymbol{\phi}}^{(i)}$, which adjusts the new search direction by accounting for the curvature of the manifold, is computed as
\begin{subequations}
\begin{equation}
\beta _{\boldsymbol{\phi }}^{\left( i \right)}=\frac{\left< \nabla _{\mathcal{R}_{\boldsymbol{\phi }}}f^{\left( i \right)},\nabla _{\mathcal{R}_{\boldsymbol{\phi }}}f^{\left( i \right)}-\nabla _{\mathcal{R}_{\boldsymbol{\phi }}}f^{\left( i-1 \right)} \right>}{ \left< \nabla _{\mathcal{R}_{\boldsymbol{\phi }}}f^{\left( i-1 \right)},\nabla _{\mathcal{R}_{\boldsymbol{\phi }}}f^{\left( i-1 \right)}  \right>},
\end{equation}
where $\left\langle \cdot, \cdot \right\rangle$ denotes the Riemannian metric (Euclidean inner product in this case). The descent direction is then updated as
\begin{equation}
\boldsymbol{\eta}^{(i)} = -\nabla _{\mathcal{R}_{\boldsymbol{\phi }}}f^{\left( i \right)}+ \beta_{\boldsymbol{\phi}}^{(i)} \mathsf{Proj}_{\boldsymbol{\phi}^{(i)}}(\boldsymbol{\eta}^{(i-1)}),
\end{equation}
\end{subequations}
where $\mathsf{Proj}_{\boldsymbol{\phi}^{(i)}}(\boldsymbol{\eta}^{(i-1)})$ is the vector transport operation that moves the previous search direction $\boldsymbol{\eta}^{(i-1)}$, also a tangent vector, from the tangent space at $\boldsymbol{\phi}^{(i-1)}$ to $\boldsymbol{\phi}^{(i)}$, ensuring appropriate mapping with the current tangent space and accounting for the curvature of the manifold. Similarly, the update coefficient and descent direction for $\boldsymbol{\theta}$ are:
\begin{subequations}
\begin{align}
&\beta _{\boldsymbol{\theta }}^{\left( i \right)}=\frac{\left< \nabla _{\mathcal{R}_{\boldsymbol{\theta }}}f^{\left( i \right)},\nabla _{\mathcal{R}_{\boldsymbol{\theta }}}f^{\left( i \right)}-\nabla _{\mathcal{R}_{\boldsymbol{\theta }}}f^{\left( i-1 \right)} \right>}{\left< \nabla _{\mathcal{R}_{\boldsymbol{\theta }}}f^{\left( i-1 \right)},\nabla _{\mathcal{R}_{\boldsymbol{\theta }}}f^{\left( i-1 \right)} \right>},
\\
&\boldsymbol{\tau}^{(i)} = -\nabla _{\mathcal{R}_{\boldsymbol{\theta }}}f^{\left( i \right)}+ \beta_{\boldsymbol{\theta}}^{(i)} \mathsf{Proj}_{\boldsymbol{\theta}^{(i)}}(\boldsymbol{\tau}^{(i-1)}).
\end{align}
\end{subequations}
For the matrix variable $\boldsymbol{X}$, the conjugate gradient update coefficient $\beta_{\boldsymbol{X}}^{(i)}$ is computed as
\begin{subequations}
\begin{equation}
\beta _{\boldsymbol{X}}^{\left( i \right)}=\frac{\left< \nabla _{\mathcal{R}_{\boldsymbol{X}}}f^{\left( i \right)},\nabla _{\mathcal{R}_{\boldsymbol{X}}}f^{\left( i \right)}-\nabla _{\mathcal{R}_{\boldsymbol{X}}}f^{\left( i-1 \right)} \right>_\mathrm{F}}{\left< \nabla _{\mathcal{R}_{\boldsymbol{X}}}f^{\left( i-1 \right)},\nabla _{\mathcal{R}_{\boldsymbol{X}}}f^{\left( i-1 \right)} \right>_\mathrm{F}},
\end{equation}
where the Riemannian metric $\left\langle \cdot, \cdot \right\rangle_\mathrm{F}$ here is the Frobenius inner product that multiplies the entries of two matrices and sums them up. The descent direction is then given by
\begin{equation}
\boldsymbol{\varGamma}^{(i)} = -\nabla _{\mathcal{R}_{\boldsymbol{X}}}f^{\left( i \right)}+ \beta_{\boldsymbol{X}}^{(i)} \boldsymbol{\varGamma}^{(i-1)},
\end{equation}
\end{subequations}
where the vector transport of the previous search direction between tangent spaces reduces to the identity mapping since the multinomial manifold has a flat geometric structure.

\subsubsection{Retraction Operation}
After updating the variables by moving along these descent directions, we map the new points back onto the manifold using retraction operations $\mathsf{Retr}(\cdot)$. This retraction ensures that the updated variables satisfy their manifold constraints, allowing the optimization to proceed within the feasible region. Specifically,
The updated variable $\boldsymbol{\phi}^{(i+1)}$ is computed as:
\begin{subequations}
\begin{align}
\boldsymbol{\phi }^{\left( i+1 \right)}&=\mathsf{Retr}_{\boldsymbol{\phi }^{\left( i \right)}}( \alpha_{\boldsymbol{\phi}} ^{\left( i \right)}\boldsymbol{\eta }^{\left( i \right)} ) 
\\
&=\left[ \frac{( \boldsymbol{\phi }^{\left( i \right)}+\alpha_{\boldsymbol{\phi}}^{\left( i \right)}\boldsymbol{\eta }^{\left( i \right)}) _m}{| ( \boldsymbol{\phi }^{\left( i \right)}+\alpha_{\boldsymbol{\phi}}^{\left( i \right)}\boldsymbol{\eta }^{\left( i \right)} ) _m |} \right] ,
\end{align}
\end{subequations}
where $\alpha_{\boldsymbol{\phi}}^{(i)}$ is the step size determined by a line search method, and the element-wise normalization ensures that $\boldsymbol{\phi}^{(i+1)}$ lies on the complex circle manifold.
Similarly, the update for $\boldsymbol{\theta}$ is
\begin{align}
\boldsymbol{\theta }^{\left( i+1 \right)}=\left[ \frac{( \boldsymbol{\theta }^{\left( i \right)}+\alpha_{\boldsymbol{\theta}}^{\left( i \right)}\boldsymbol{\tau }^{\left( i \right)})_n}{| ( \boldsymbol{\theta }^{\left( i \right)}+\alpha_{\boldsymbol{\theta}}^{\left( i \right)}\boldsymbol{\tau }^{\left( i \right)} )_n |} \right].
\end{align}
For $\boldsymbol{X}$, the retraction involves projecting the updated $\boldsymbol{X}$ back onto the multinomial manifold:
\begin{subequations}
\begin{align}
\boldsymbol{X}^{(t+1)}&=\mathsf{Retr}_{\boldsymbol{X}^{\left( i \right)}}( \alpha_{\boldsymbol{X}}^{\left( i \right)}\boldsymbol{\varGamma}^{\left( i \right)} ) 
\\
& = \Pi_{\mathcal{R}_{\boldsymbol{X}}}\left( \boldsymbol{X}^{(i)} + \alpha_{\boldsymbol{X}}^{\left( i \right)} \boldsymbol{\varGamma}_{\boldsymbol{X}}^{(i)} \right),
\end{align}
\end{subequations}
where $\Pi_{\mathcal{R}_{\boldsymbol{X}}}$ denotes the projection onto the multinomial manifold, which can be performed using the algorithm in [40].

\vspace{-12pt}
\subsection{Overall Algorithm}
Algorithm 2 summarizes the complete manifold optimization procedure, initialized with feasible random points within each variable's respective manifolds, and iteratively updating all variables using the RCG method until convergence. At convergence, thresholding techniques are applied to the scheduling matrix to recover binary decisions, while manifold structures preserve unit-modulus constraints. The RCG method on these manifolds guarantees convergence to a stationary point under standard regularity conditions (e.g., Lipschitz continuity of gradients, appropriate line-search steps) [39].

Let $I_{\mathrm{RCG}}$ denote the number of RCG iterations. In each iteration, the main cost comes from computing the objective and Riemannian gradients w.r.t. $\boldsymbol{\phi}$, $\boldsymbol{\theta}$, and the scheduling matrix. For each user–pattern pair $(k, u)$, evaluating $\gamma_{k,u} = |{\bf v}_u^H \boldsymbol{\Xi}_k {\bf v}_u|$ incurs $\mathcal{O}(M)$ and $\mathcal{O}(N)$ operations for gradients w.r.t. $\boldsymbol{\phi}$ and $\boldsymbol{\theta}$, respectively, assuming that $\boldsymbol{\Xi}_k$ is precomputed. The sum of all pairs yields $\mathcal{O}(KU(M+N))$ per iteration. Retractions contribute lower-order terms: $\mathcal{O}(KU \log U)$ for the simplex projection and $\mathcal{O}(M+N)$ for the complex circle projection. Hence, the complexity per iteration is dominated by $\mathcal{O}(KU(M+N))$, leading to a total complexity of $\mathcal{O}\left(I_{\mathrm{RCG}}\,KU\,(M+N)\right)$, which is significantly lower than the $\mathcal{O}((KU)^3)$ complexity of interior-points based BCD methods.

\renewcommand{\algorithmicrequire}{\textbf{Input:}}
\renewcommand{\algorithmicensure}{\textbf{Output:}}
\begin{algorithm}[!t]
\caption{Manifold Optimization Algorithm for (P2)}
\label{alg:general_solution}
\begin{algorithmic}[1]
\REQUIRE $N_c, N_r, M_c, M_r, K, \{\boldsymbol{S}_u\}$, $\{\boldsymbol{e}_u\}$, and $\{\boldsymbol{\varXi}_k\}$
\ENSURE $\boldsymbol{\phi}^\star, \boldsymbol{\theta}^\star,\boldsymbol{X}^\star$
\STATE Initialize $\boldsymbol{\phi}^{(0)}$, $\boldsymbol{\theta}^{(0)}$, and $\boldsymbol{X}^{(0)}$ within their respective manifolds; Set initial descent directions $\boldsymbol{\eta}^{(0)},\boldsymbol{\tau}^{(0)}$, and $\boldsymbol{\varGamma}^{(0)}$; Set iteration index $i = 0$.
        \WHILE{the norm of Riemannian gradient $\big(\lVert \nabla _{\mathcal{R}_{\boldsymbol{\phi }}}f^{(i)} \rVert _{2}^{2}+\lVert \nabla _{\mathcal{R}_{\boldsymbol{\theta }}}f^{(i)}\rVert _{2}^{2}+\lVert \nabla _{\mathcal{R}_{\boldsymbol{X}}}f^{(i)} \rVert _{\text{F}}^{2}\big)^{\frac{1}{2}}$ is above the threshold}
            \STATE Set iteration index $i = i + 1$.
            \STATE Calculate step size $\alpha_{\boldsymbol{\phi}}^{(i-1)}$, $\alpha_{\boldsymbol{\theta}}^{(i-1)}$, and $\alpha_{\boldsymbol{X}}^{(i-1)}$ using backtracking algorithms [39].
            \STATE Update $\boldsymbol{\phi}^{(i)}$ using (36a), (37), (40), and (43).
            \STATE Update $\boldsymbol{\theta}^{(i)}$ using (36b), (38), (41), and (44).
            \STATE Update $\boldsymbol{X}^{(i)}$ using (36c), (39), (42), and (45).
        \ENDWHILE
\RETURN $\boldsymbol{\phi}^\star=\boldsymbol{\phi}^{(i)}$, $\boldsymbol{\theta}^\star=\boldsymbol{\theta}^{(i)}$, and $\boldsymbol{X}^\star=\boldsymbol{X}^{(i)}$.
\end{algorithmic}
\end{algorithm}

\vspace{-12pt}
\section{Element-wise Tunable Position}

In this section, we investigate the theoretical performance limits of the MIS system by considering a more flexible design, where each element of the movable MS2 can be independently repositioned to optimally overlap with the elements of MS1. This idealized configuration provides an upper bound on system performance, thereby revealing the fundamental gap between static phase shift with mechanical reconfigurability (not limited to the block-wise control architecture in previous sections) and the capabilities of a fully dynamic RIS.

In contrast to the original MIS architecture, which restricts MS2 movement to block-wise shifting as a whole, the present approach treats the selection matrices ${\boldsymbol{S}_u}$ and the corresponding padding vectors ${\boldsymbol{e}_u}$ in (2) as fully tunable optimization variables, rather than confining them to a predefined set of admissible configurations (e.g, moving MS2 with several units as a whole and then mapping them onto MS1). This high degree of flexibility enables, under a TDMA protocol, each user $k$ to be served by a user-specific optimized overlap pattern, denoted by $\boldsymbol{S}_k$. Accordingly, the resulting optimization problem can be formulated as
\begin{subequations}
\begin{align}
\text{(P6)}:&\underset{\boldsymbol{\phi },\boldsymbol{\theta },\left\{ \boldsymbol{S}_k \right\} ,\left\{ \boldsymbol{e}_k \right\}}{\max}\sum_{k\in \mathcal{K}}{\log _2\left( 1+\boldsymbol{v}_{k}^{H}\boldsymbol{\varXi }_k\boldsymbol{v}_{k}^{} \right)}
\\
\text{s.t.} \,\,& \boldsymbol{v}_{k}^{}=\boldsymbol{\phi }\odot \left( \boldsymbol{S}_k\boldsymbol{\theta }+\boldsymbol{e}_k \right), 
\\
&\text{(2a), (2b), (10e), (10f)}. \nonumber 
\end{align}
\end{subequations}
Here, constraints (2a) and (2b), which capture the physical overlapping conditions between the MS layers, remain implicit. We explicitly restate these constraints as follows
\begin{subequations}
\begin{align}
&\left[ \boldsymbol{e}_k \right] _m=1-\sum_{n\in \mathcal{N}}{\left[ \boldsymbol{S}_k \right] _{m,n}}, \forall k\in \mathcal{K}, \forall m\in \mathcal{M},
\\
&\sum_{m\in \mathcal{M}}{\left[ \boldsymbol{S}_k \right] _{m,n}}=1,\forall k\in \mathcal{K},\forall n\in \mathcal{N},
\\
&\sum_{n\in \mathcal{N}}{\left[ \boldsymbol{S}_k \right] _{m,n}}\le 1,\forall k\in \mathcal{K},\forall m\in \mathcal{M},
\\
&\left[ \boldsymbol{S}_k \right] _{m,n}\in \left\{ 0,1 \right\} ,\forall n\in \mathcal{N},\forall m\in \mathcal{M}.
\end{align}
\end{subequations}
Constraint (47a) directly relates the padding vector $\boldsymbol{e}_k$ to $\boldsymbol{S}_k$, allowing $\boldsymbol{e}_k$ to be eliminated from the optimization variables. (47b) ensures that each element of MS2 overlaps exactly one element of MS1, while (47c) guarantees that no more than one element of MS2 is placed at any given MS1 position.

Consistent with the approach in Section V, we employ manifold optimization to jointly optimize all variables. Apart from the standard manifold constraint on MIS phase shifts, constraints (47b) and (47d) motivate the construction of a multinomial manifold by relaxing the binary constraints as follows
\begin{align}
\sum_{m\in \mathcal{M}}{\left[ \boldsymbol{S}_k \right] _{m,n}}=1, \left[ \boldsymbol{S}_k \right] _{m,n}\ge 0, \forall k,m,n.
\end{align}
Unlike earlier scenarios, this continuous relaxation does not inherently guarantee binary solutions during optimization. Additionally, constraint (47c) is not directly enforced by this relaxation. To address these challenges, we introduce two quadratic penalty terms
\begin{subequations}
\begin{align}
&p\left( \left\{ \boldsymbol{S}_k \right\} \right) =\frac{1}{N}\left( \left[ \boldsymbol{S}_k \right] _{m,n}-\frac{1}{2} \right) ^2
\\
&q\left( \left\{ \boldsymbol{S}_k \right\} \right) =\Big( \max \Big\{ \sum_{n\in \mathcal{N}}{\left[ \boldsymbol{S}_k \right] _{m,n}},1 \Big\} -1 \Big) ^2
\end{align}
\end{subequations}
Subsequently, following the procedures defined in (31)-(34), which specify the product manifold and tangent space, problem (P6) can be reformulated as an unconstrained manifold optimization problem (P7) on the top of the next page,
\begin{figure*}
\begin{align}
&\!\text{(P7)}:\!\underset{\boldsymbol{\phi },\boldsymbol{\theta },\left\{ \boldsymbol{S}_k \right\} \in \mathcal{R}_{\left( \boldsymbol{\phi },\boldsymbol{\theta },\left\{ \boldsymbol{S}_k \right\} \right)}}{\max} \!\! F(\boldsymbol{\phi },\boldsymbol{\theta },\{ \boldsymbol{S}_k \}) \!=\!\sum_{k\in \mathcal{K}}{\log _2\left( 1+\boldsymbol{v}_{k}^{H}\boldsymbol{\varXi }_k\boldsymbol{v}_{k}^{} \right)}\!+\!\sum_{k\in \mathcal{K}}\!{\sum_{m\in \mathcal{M}}\!{\sum_{n\in \mathcal{N}}{p_{m,n}\left( \left\{ \boldsymbol{S}_k \right\} \right)}}}\!-\!\!\sum_{k\in \mathcal{K}}\!{\sum_{m\in \mathcal{M}}\!{q_m\left( \left\{ \boldsymbol{S}_k \right\} \right)}}.\!\!
\\
& \overline{\ \ \ \ \ \ \ \ \ \ \ \ \ \ \ \ \ \ \ \ \ \ \ \ \ \ \ \ \ \ \ \ \ \ \ \ \ \ \ \ \ \ \ \ \ \ \ \ \ \ \ \ \ \ \ \ \ \ \ \ \ \ \ \ \ \ \ \ \ \ \ \ \ \ \ \ \ \ \ \ \ \ \ \ \ \ \ \ \ \ \ \ \ \ \ \ \ \ \ \ \ \ \ \ \ \ \ \ \ \ \ \ \ \ \ \ \ \ \ \ \ \ \ \ \ \ \ \ \ \ \ \ \ \ \ \ \ \ \ \ \ \ \ \ \ } \nonumber
\end{align}
\vspace{-36pt}
\end{figure*}
with the product manifold structure defined as
\begin{align}
\mathcal{R}_{\left( \boldsymbol{\phi },\boldsymbol{\theta },\left\{ \boldsymbol{S}_k \right\} \right)}=\mathcal{R}_{\boldsymbol{\phi }}\times \mathcal{R}_{\boldsymbol{\theta }}\times \prod_{k\in \mathcal{K}}{\mathcal{R}_{\boldsymbol{S}_k}},
\end{align}
where $\mathcal{R}_{\boldsymbol{\phi }}$ and $\mathcal{R}_{\boldsymbol{\theta }}$ denote complex circle manifolds, and $\mathcal{R}_{\boldsymbol{S}_k}$represents the multinomial manifold for each user $k$
\begin{align}
    &\mathcal{R}_{\boldsymbol{S}_k} \! =\! \Big\{ \!\boldsymbol{S}_k \!\in \!\mathbb{R}^{M \times N}\!:\! \sum_{m\in \mathcal{M}} \![\boldsymbol{S}_k]_{m,n}\! =\!1, [\boldsymbol{S}_k]_{m,n} \!>\!0, \!\forall m, \!\forall n \Big\}.
\end{align}

We again apply the RCG method to solve (P7). The Euclidean gradients of the objective function w.r.t to each variable are analytically derived in (55) and projected onto their corresponding tangent spaces. The descent directions are determined using the Polak–Ribiere method, and manifold-specific retraction operations are performed as outlined in Section VI. The overall optimization procedure is summarized in Algorithm 3. 

In each iteration, the primary computational costs involve evaluating the gradient w.r.t. the selection matrices $\{\boldsymbol{S}_k\}$, and phase vectors $\boldsymbol{\phi}$ and $\boldsymbol{\theta}$. For each user, constructing the composite vector $\boldsymbol{v}_k$ requires $\mathcal{O}(MN)$ operations, and multiplying it by the $M \times M$ matrix $\boldsymbol{\Xi}_k$ incurs $\mathcal{O}(M^2)$ complexity. Additional element-wise and penalty computations remain linear, i.e., $\mathcal{O}(MN)$. Thus, the per-iteration gradient evaluation has a complexity of  $\mathcal{O}(K(MN+M^2))$. The computational costs of retractions and projections are negligible in comparison. As a result, the overall complexity of the algorithm is $\mathcal{O}(I_{\mathrm{RCG}}\,K(MN+M^2))$, where $I_{\mathrm{RCG}}$ denotes the number of RCG iterations.
\begin{figure*}
\begin{subequations}
\begin{align}
& \qquad \nabla _{\boldsymbol{\phi }}F=\frac{2}{\ln 2}\sum_{k=1}^K{\frac{1}{1+\boldsymbol{v}_{k}^{H}\boldsymbol{\varXi }_k\boldsymbol{v}_{k}^{}}\mathrm{diag}\left( \boldsymbol{S}_k\boldsymbol{\theta }^*+\boldsymbol{e}_k \right) \boldsymbol{\varXi }_k\mathrm{diag}\left( \boldsymbol{S}_k\boldsymbol{\theta }+\boldsymbol{e}_k \right) \boldsymbol{\phi }},
\\
& \qquad \nabla _{\boldsymbol{\theta }}F=\frac{2}{\ln 2}\sum_{k=1}^K{\frac{1}{1+\boldsymbol{v}_{k}^{H}\boldsymbol{\varXi }_k\boldsymbol{v}_{k}^{}}\boldsymbol{S}_{k}^{T}\mathrm{diag}\left( \boldsymbol{\phi }^* \right) \boldsymbol{\varXi }_k\mathrm{diag}\left( \boldsymbol{\phi } \right) \left( \boldsymbol{S}_k\boldsymbol{\theta }+\boldsymbol{e}_k \right)},
\\
& \qquad \nabla _{\boldsymbol{S}_k}F=\frac{2}{\ln 2}\frac{1}{1+\boldsymbol{v}_{k}^{H}\boldsymbol{\varXi }_k\boldsymbol{v}_{k}^{}}\Re\big\{ \mathrm{diag}\left( \boldsymbol{\phi }^* \right) \boldsymbol{\varXi }_k\mathrm{diag}\left( \boldsymbol{\phi } \right) \left( \boldsymbol{S}_k\boldsymbol{\theta }+\boldsymbol{e}_k \right) \left( \boldsymbol{\theta }-\mathbf{1}_N \right) ^H \big\} -2\big( \boldsymbol{S}_k-\frac{1}{2}\mathbf{1}_{M\times N} \big) -2\boldsymbol{\varDelta }_k,
\\
&\qquad\left[ \boldsymbol{\varDelta }_k \right] _{m,:}=\Big[ \max \Big\{ \sum_{n\in \mathcal{N}}{\left[ \boldsymbol{S}_k \right] _{m,n}}-1,0 \Big\} ,\dots ,\max \Big\{ \sum_{n\in \mathcal{N}}{\left[ \boldsymbol{S}_k \right] _{m,n}}-1,0 \Big\} \Big]. 
\\
& \overline{\ \ \ \ \ \ \ \ \ \ \ \ \ \ \ \ \ \ \ \ \ \ \ \ \ \ \ \ \ \ \ \ \ \ \ \ \ \ \ \ \ \ \ \ \ \ \ \ \ \ \ \ \ \ \ \ \ \ \ \ \ \ \ \ \ \ \ \ \ \ \ \ \ \ \ \ \ \ \ \ \ \ \ \ \ \ \ \ \ \ \ \ \ \ \ \ \ \ \ \ \ \ \ \ \ \ \ \ \ \ \ \ \ \ \ \ \ \ \ \ \ \ \ \ \ \ \ \ \ \ \ \ \ \ \ \ \ \ \ \ \ \ \ \ \ } \nonumber
\end{align}
\vspace{-36pt}
\end{subequations}
\end{figure*}

\begin{algorithm}[!t]
\caption{Penalty-Assisted Manifold Optimization for (P6)}
\label{alg:joint_manopt}
\begin{algorithmic}[1]
\REQUIRE $N_c, N_r, M_c, M_r, K,$ and $\{\boldsymbol{\varXi}_k\}$
\ENSURE $\boldsymbol\phi^\star,\;\boldsymbol\theta^\star,\;\{\boldsymbol{S}_k^\star\}$
\STATE Initialize $\boldsymbol{\phi}^{(0)}$, $\boldsymbol{\theta}^{(0)}$, and $\{\boldsymbol{S}_k^{(0)}\}$ in their respective manifolds, descent directions, and iteration index $i=0$.
\WHILE{$\big( \!\lVert \nabla \!_{\mathcal{R}_{\boldsymbol{\phi }}}\!F^{\left( i \right)} \rVert _{2}^{2} \! + \!\lVert \nabla \!_{\mathcal{R}_{\boldsymbol{\theta }}}\!F^{\left( i \right)} \rVert _{2}^{2}\!+\!\sum_{k\in \mathcal{K}}{\lVert \nabla \!_{\mathcal{R}_{\boldsymbol{S}_k}}\!F^{\left( i \right)} \rVert _{\text{F}}^{2}}\!\big) ^{\frac{1}{2}}$ is above the threshold}
    \STATE Set iteration index $i = i + 1$.
    \STATE Calculate step size using backtracking algorithms.
    \STATE Update $\boldsymbol{\phi}^{(i)}$, $\boldsymbol{\theta}^{(i)}$, and $\{\boldsymbol{S}^{(i)}_k\}$ based on (53) and RCG operations.
\ENDWHILE
\RETURN $\boldsymbol{\phi}^\star=\boldsymbol{\phi}^{(i)}$, $\boldsymbol{\theta}^\star=\boldsymbol{\theta}^{(i)}$, and $\{\boldsymbol{S}^\star_k\}=\{\boldsymbol{S}^{(i)}_k\}$ after binary thresholding.
\end{algorithmic}
\end{algorithm}

\vspace{-16pt}
\section{Numerical Results}
\vspace{-2pt}
This section provides numerical results evaluating the performance of our proposed MIS-assisted schemes: (1) minimum data rate maximization, (2) total throughput maximization, and (3) element-wise movable MS2 case. In the simulations, users are placed around the MIS at a fixed elevation angle $\psi_k=-\frac{\pi}{4},\forall k$ with azimuth angles uniformly distributed within $\psi_k \in [0,\frac{\pi}{3}]$ at equal distances from the MIS. The large-scale path loss and noise power are combined, resulting in a reference SNR of $\gamma_{\text{ref}}=0.05$ for all users. Unless otherwise specified, we set the small surface to $N_r=N_c=4$, the Rician factor $\kappa=10$ dB for all channels involved, the number of BS antennas $L = 4$, the transmit power $P = 30$ dBm, and the total communication time $T=100s$ by default.

\setlength{\abovecaptionskip}{5pt}
\begin{figure}[t]
\centering
\includegraphics[width=2.8in]{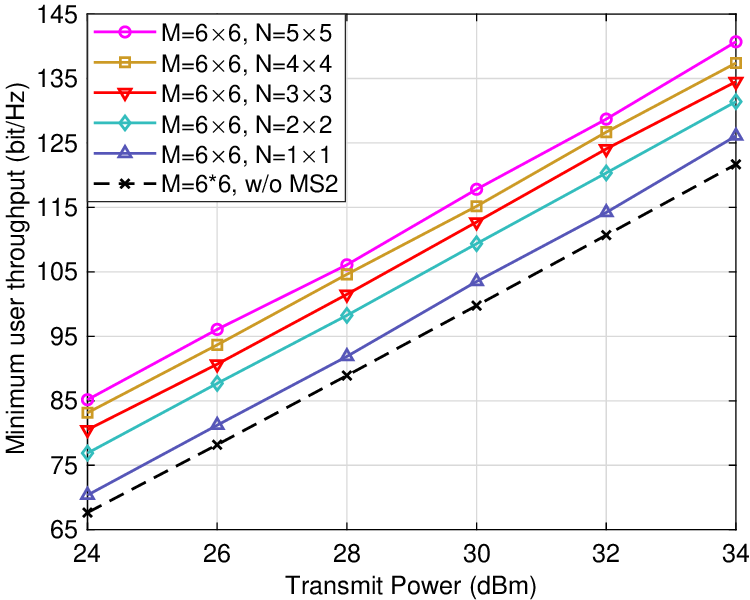}
\captionsetup{font=small}
\caption{Minimum user throughput vs. transmit power $P$ with different MS2 configurations.} 
\vspace{-12pt}
\end{figure}
\begin{figure}[t]
\centering
\includegraphics[width=2.8in]{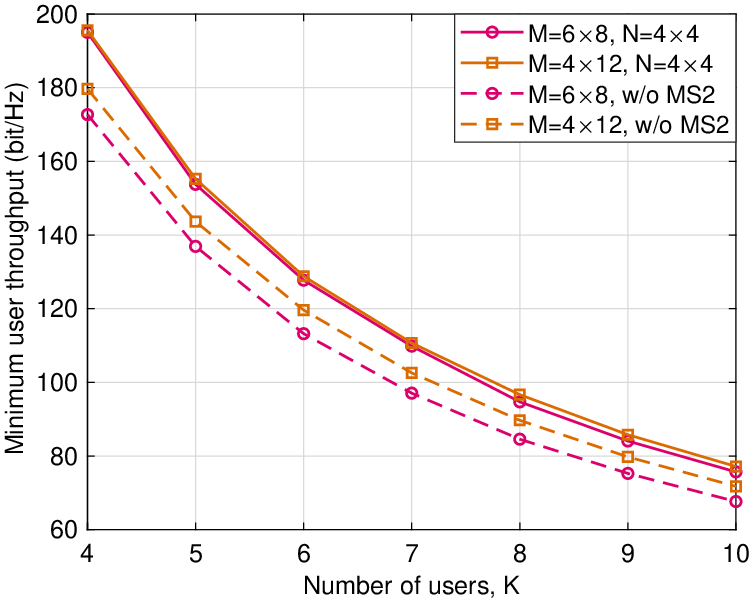}
\captionsetup{font=small}
\caption{Minimum user throughput vs. the number of users $K$.} 
\vspace{-9pt}
\end{figure}

\vspace{-13pt}
\subsection{Minimum Data Rate Maximization (Algorithm 1)}

\vspace{-1pt}
\subsubsection{Impact of transmit power and MS2 size}
Fig. 3 illustrates the minimum data rate versus transmit power ($P=24$–$34$ dBm) for varying MS2 sizes (from $1\times1$ to $5\times5$). The minimum rate increases logarithmically with transmit power, and the larger MS2 sizes consistently provide higher rates. Even the smallest MS2 configuration ($1\times1$) notably surpasses the single-layer static surface baseline, confirming the effectiveness of the MIS architecture in improving performance.

\subsubsection{Impact of number of users}
Fig. 4 evaluates how the minimum data rate changes as the number of users $K$ increases from $4$ to $10$, with different MS1 configurations and a fixed total number of elements. The minimum rate stably decreases with increasing users, primarily due to TDMA scheduling limitations. Nevertheless, MIS configurations maintain a clear advantage, delivering around 10–15 bit/Hz higher minimum data rate than the baseline system without MS2, demonstrating robust performance in densely loaded scenarios. Furthermore, varying the aspect ratio of MS1 only has a minor influence on performance.

\begin{figure}[t]
\centering
\includegraphics[width=2.8in]{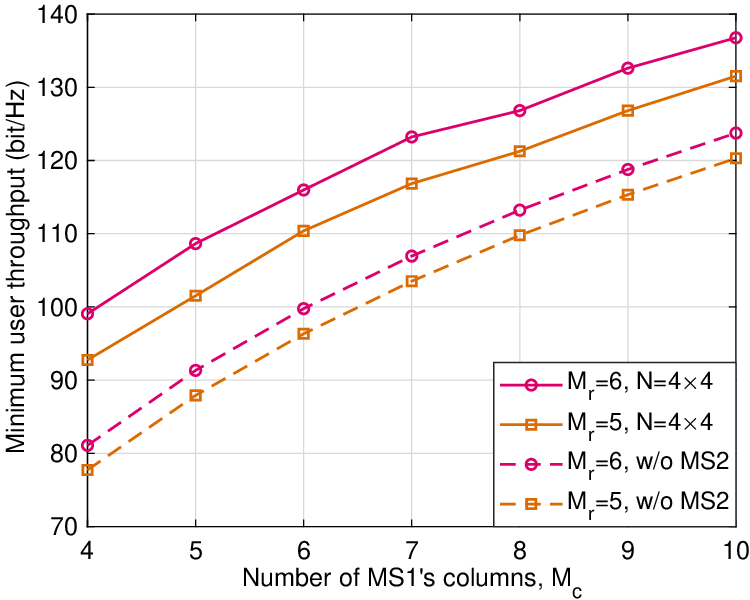}
\captionsetup{font=small}
\caption{Minimum user throughput vs. MS1 array size $M$.} 
\vspace{-9pt}
\end{figure}
\begin{figure}[t]
\centering
\includegraphics[width=2.8in]{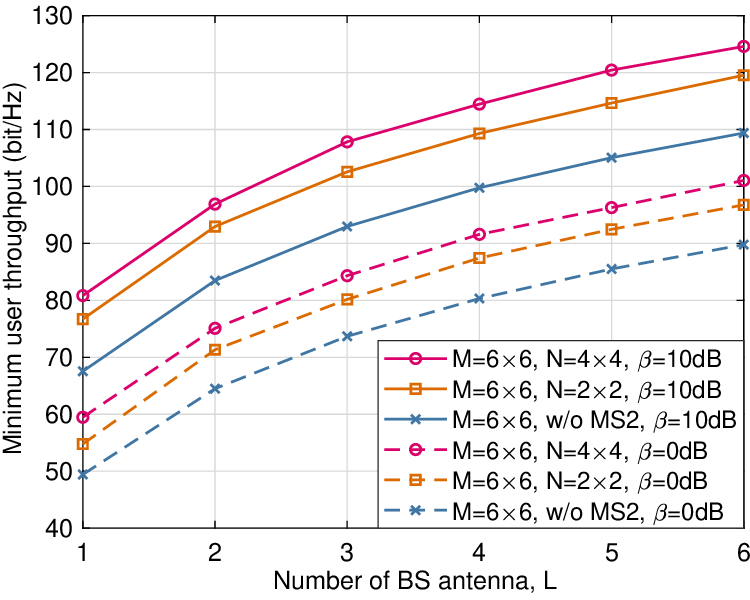}
\captionsetup{font=small}
\caption{Minimum user throughput vs. the number of BS antennas $L$ with different Rician factors.} 
\vspace{-9pt}
\end{figure} 

\subsubsection{Impact of MS1 array size}
Fig. 5 compares the minimum data rate performance for two MS1 array configurations ($M_r = 6$ and $M_r = 5$) as the number of columns $M_c$ varies from 4 to 10. Across all cases, the MIS consistently outperforms the single-layer static surface baseline, achieving performance gains of up to 20 bit/Hz. The results reveal that increasing $M_c$ yields diminishing returns, as the performance curve flattens for larger apertures. Notably, introducing MS2 provides additional gains even when the overall aperture size is kept constant, confirming the effectiveness of the MIS working mechanism. This highlights that the benefits brought by movable surfaces are comparable to, and in some cases can complement, the impact of aperture size in achieving higher data rates.

\subsubsection{Impact of BS antennas and Rician factors}
Fig. 6 analyzes how the number of BS antennas ($L=1$–$6$) and MS2 size ($4\times4$ vs. $2\times2$) influence the minimum rate under different Rician factors ($\kappa=10$ dB and $\kappa=0$ dB). Performance improves steadily with more antennas and is notably higher in stronger LoS conditions, attributed to the MIS static phase shift and beam pattern scheduling design based on statistical CSI. Additionally, a larger MS2 configuration ($4\times4$) provides an extra 5–10 bit/Hz rate increase compared to the smaller configuration.

\vspace{-12pt}
\subsection{Total Throughput Maximization (Algorithm 2)}

\begin{figure}[t]
\centering
\includegraphics[width=2.8in]{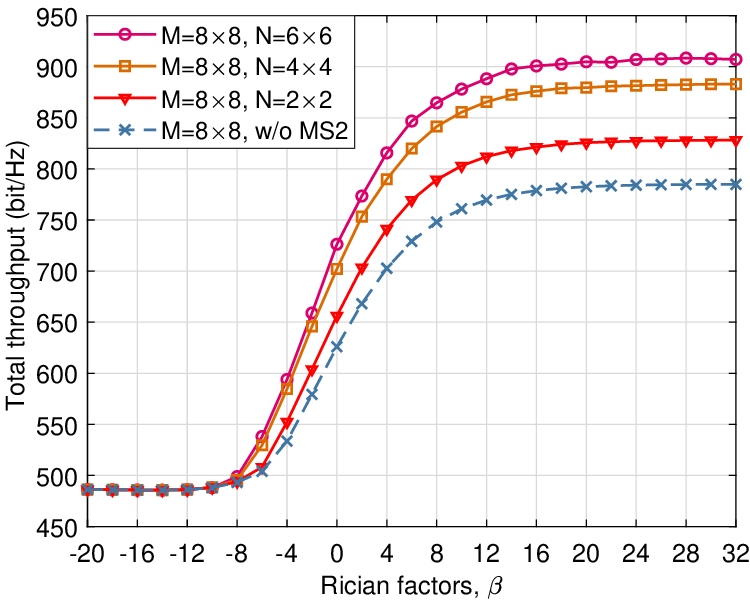}
\captionsetup{font=small}
\caption{Total throughput vs. Rician factors with varying MS2 sizes.} 
\vspace{-9pt}
\end{figure}
\begin{figure}[t]
\centering
\includegraphics[width=2.8in]{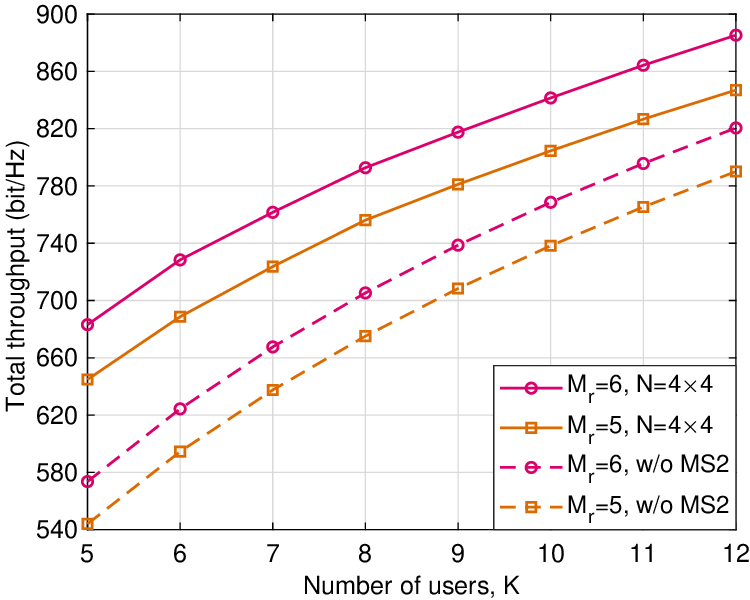}
\captionsetup{font=small}
\caption{Total throughput vs. MS1 array size $M$.} 
\vspace{-6pt}
\end{figure}

\subsubsection{Impact of Rician factor on throughput}
Fig. 7 shows the effect of varying the Rician factor ($\kappa$) on total throughput performance. Throughput significantly improves as the Rician factor increases, particularly for larger MS2 configurations ($M = 8 \times 8, N = 6 \times 6$). Relative to the single-layer static surface baseline, the MIS architecture accelerates throughput growth and achieves higher performance ceilings under strong LoS scenarios. As $\kappa$ becomes sufficiently large, the throughput saturates, which is consistent with theoretical expectations that stronger LoS components improve beamforming accuracy and reliability.

\subsubsection{Impact of MS1 array size on throughput}
Fig. 8 illustrates throughput improvements as the MS1 array size increases with fixed MS2 size. Similar insights from Fig. 3 are observed here. In addition, as the MS1 array size increases, the throughput improvement decelerates, and the trend gradually falls to the performance of a single-layer static surface.

\begin{figure}[t]
\centering
\includegraphics[width=2.8in]{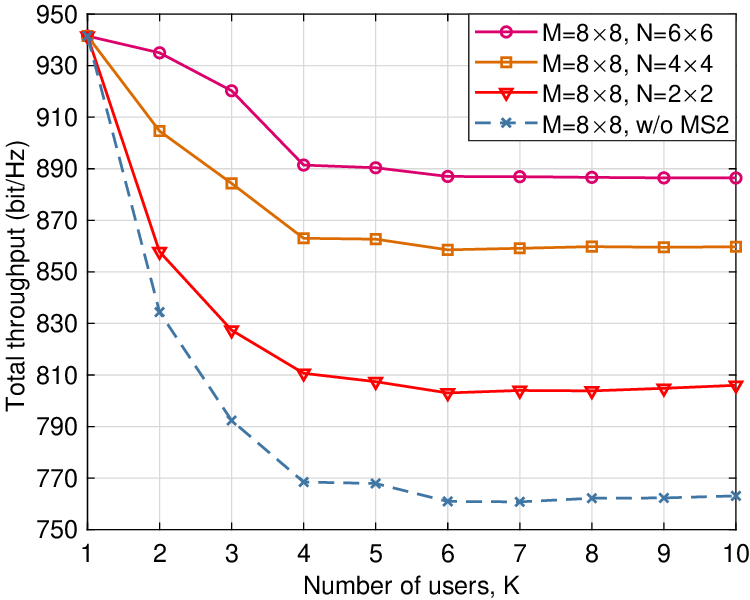}
\captionsetup{font=small}
\caption{Total throughput vs. the number of users $K$.} 
\vspace{-9pt}
\end{figure}
\begin{figure}[t]
\centering
\includegraphics[width=2.8in]{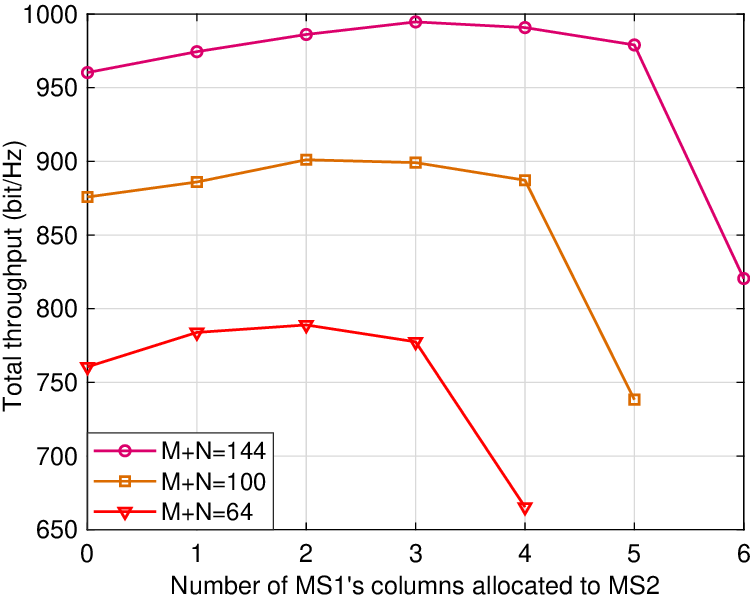}
\captionsetup{font=small}
\caption{Total throughput of the MIS scheme under different strategies of element allocation between MS 1 and MS 2} 
\vspace{-6pt}
\end{figure}

\subsubsection{Impact of number of users on throughput}
Fig. 9 shows that as the number of users $K$ increases, the total throughput decreases, because the system needs to share resources (beam patterns and time slots) between users, which reduces the available resources per user. However, configurations with larger MS2 arrays ($M = 8 \times 8, N = 6 \times 6$) better maintain throughput levels under higher user densities. While a smaller MS2 enables the generation of more beam patterns, only a sufficiently large MS2 can ensure that these patterns are adequately distinct. Although increasing MS2 size may reduce the number of available patterns due to physical restrictions, it improves the diversity and separation of the patterns. This trade-off reveals that, despite fewer available patterns, increased pattern distinctiveness leads to more efficient multiuser support and superior overall system performance.

\subsubsection{Impact of element allocation between MS1 and MS2 on throughput}
Fig. 10 examines the optimal allocation of elements between MS1 and MS2 with a fixed total number of elements in the MIS. A configuration with zero elements in MS2 reduces the system to a single-layer static surface baseline for a fair comparison. We progressively transfer rows from MS1 to MS2 by fixing the columns of MS1 and the rows of MS2 while dividing the rows from MS1 to MS2, ensuring that the rows of MS2 are equal to half of the rows of MS1 (e.g., transition from MS1: $8 \times 8$ and MS2: $4 \times 0$ to MS1: $4 \times 8$ and MS2: $4 \times 8$ when $M+N=64$). It is shown that moderate allocations of elements to MS2 maximize throughput, while excessive allocations to MS2 degrade performance due to insufficient aperture gain on MS1, revealing a trade-off and its optimal allocation strategy for balancing beamforming flexibility and aperture gain.

\vspace{-6pt}
\subsection{Element-wise movable MS2 optimization (Algorithm 3)}

\begin{figure}[t]
\centering
\includegraphics[width=2.8in]{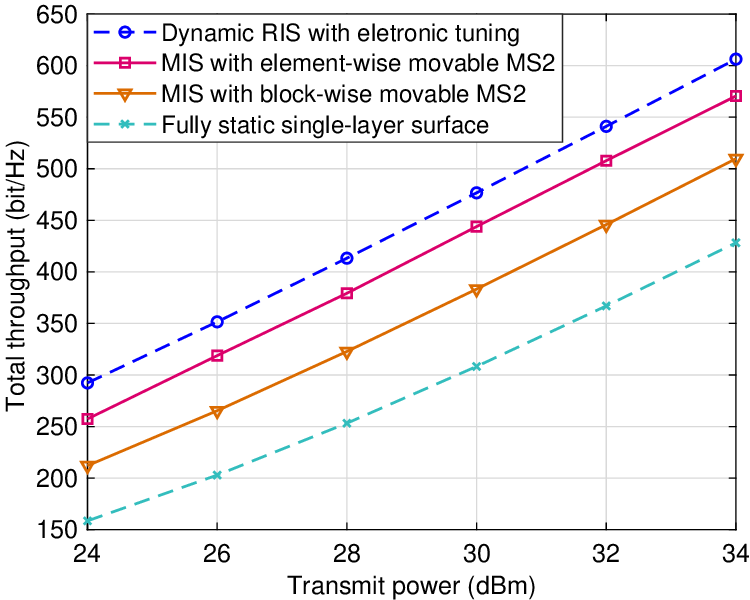}
\captionsetup{font=small}
\caption{Element-wise vs. Block-level MS2 Mobility.} 
\vspace{-9pt}
\end{figure}

Fig. 11 compares the total throughput among four scenarios: element-wise movable MS2 (Algorithm 3), block-level movable MS2 (Algorithm 1), single-layer static surface, and dynamic RIS, varying transmit power ($24$–$34$ dBm), with MS1 size $5\times5$ and MS2 size $3\times3$. The element-wise mechanical reconfiguration scheme significantly outperforms the block-level shifting approach by approximately 65 bit/Hz and the single-layer static surface by around 140 bit/Hz at the maximum power. Importantly, the element-wise movable MS2 scheme can recover approximately 60\% of the throughput gap between block-level MIS and fully dynamic RIS, owing to the flexibility of element-level repositioning. This result demonstrates the theoretical performance ceiling achievable by the MIS architecture when limited to static phase shifts.

\vspace{-12pt}
\subsection{Performance Gap of the Jensen Upper Bound}
\begin{figure}[t]
\centering
\includegraphics[width=2.8in]{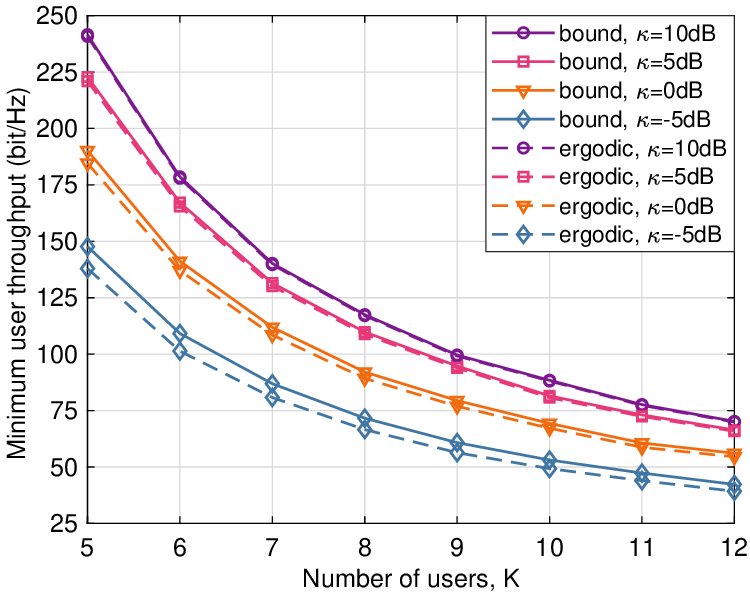}
\captionsetup{font=small}
\caption{Comparison of the upper bound and true ergodic value of minimum user throughput vs. number of users $K$.} 
\vspace{-9pt}
\end{figure}
\begin{figure}[t]
\centering
\includegraphics[width=2.8in]{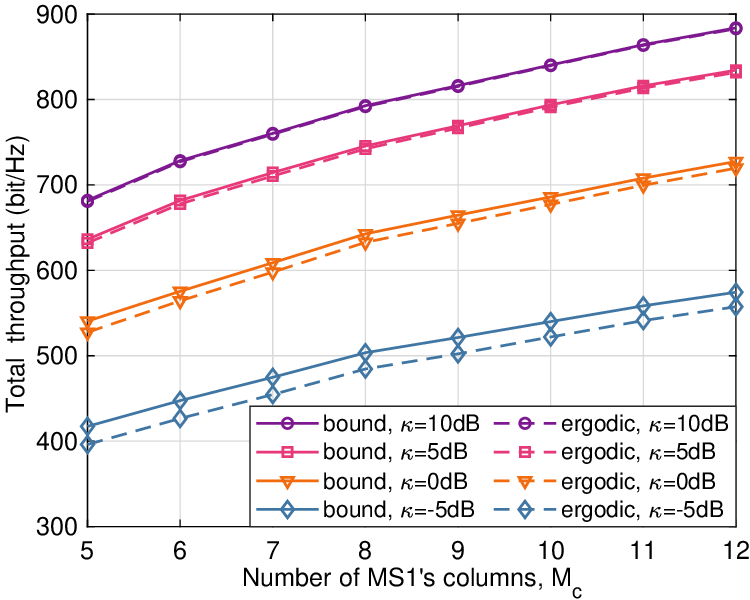}
\captionsetup{font=small}
\caption{Comparison of the upper bound and true ergodic value of total throughput vs. number of MS1 columns $M_c$.} 
\vspace{-6pt}
\end{figure}

In Fig. 12, we set $M=6\times6$ and $N=4\times4$ with different Rician factors and number of users. For all Rician factors ($\kappa\in{-5,0,5,10},\mathrm{dB}$), the minimum user throughput decreases with $K$, as expected from stronger multiuser competition. In every pair of curves, the upper ``bound" lies slightly above the true ``ergodic" curve with a small gap. The gap shrinks as $\kappa$ increases with a stronger LoS component, indicating a tighter concentration of $\gamma_{k,u}$) around its mean, resulting in a tighter Jensen bound. In particular, the optimizing beam-pattern schedule derived from the bound shows a trend almost identical to the true ergodic objective. In Fig. 13, we set $K=6$, $M_r=6$, $N=4\times4$ with different Rician factors and number of MS1 columns $M_c$. The total throughput increases monotonically with aperture size for all $kappa$. The “bound” and “ergodic” curves are again almost parallel, with a small, stable offset that decreases as $\kappa$ increases. Thus, the configuration that maximizes the bound also maximizes the realized sum throughput for a given system configuration. 

Across operating points, the bound only modestly overestimates the achieved ergodic rate, especially in moderate/high $\kappa$. More importantly, the same scaling trend between the upper bound and actual ergodic rate holds when optimizing the former objective based on Jensen's equality, retaining the sub-optimal choices on beam pattern scheduling and static phase shifts under a more tractable problem formulation. In summary, the empirical evidence supports using the Jensen bound as the optimization objective: it preserves the correct trends with its achieved ergodic rate counterpart, delivers a feasible, high-quality sub-optimal solution with limited performance gap in practice, and enables a tractable, differentiable objective for our BCD-SCA or manifold-based solver with substantial computational savings.

\vspace{-12pt}
\subsection{Performance Comparison with the Benchmark Algorithms}
\begin{figure}[t]
\centering
\includegraphics[width=2.8in]{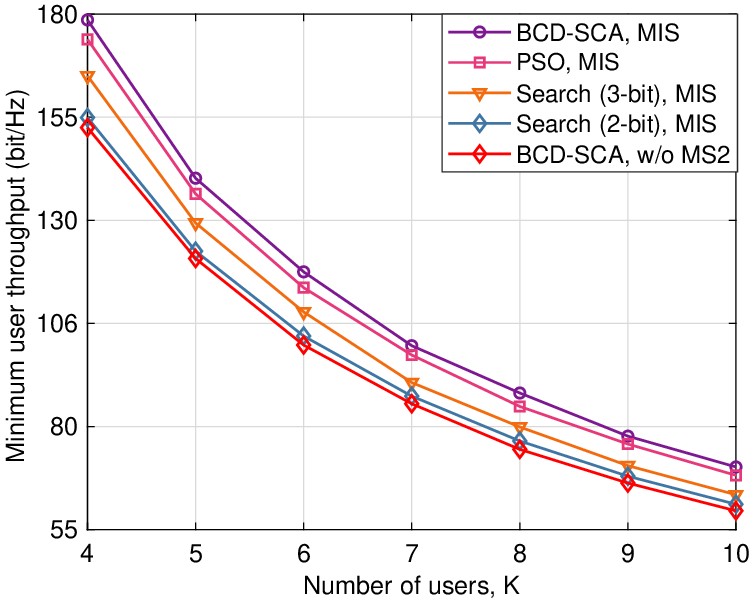}
\captionsetup{font=small}
\caption{Comparison of benchmarks with the BCD-SCA algorithm of minimum user throughput vs. number of users $K$.} 
\vspace{-6pt}
\end{figure}
\begin{figure}[t]
\centering
\includegraphics[width=2.8in]{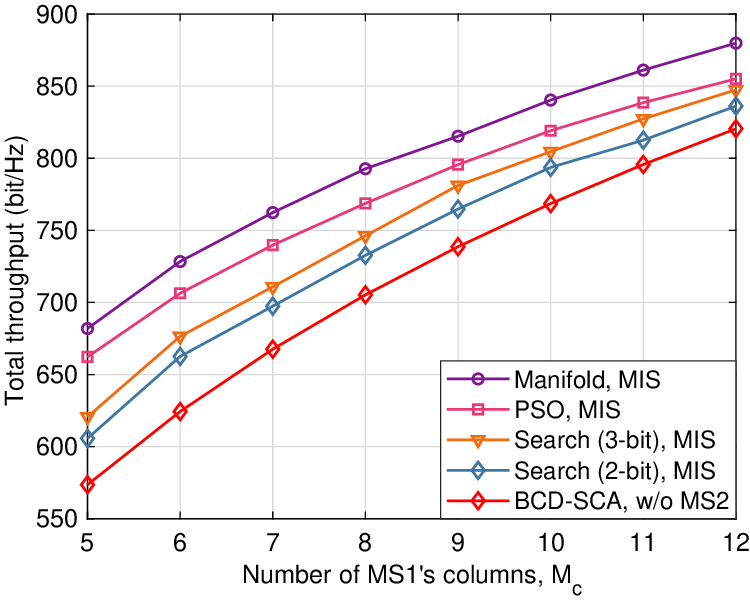}
\captionsetup{font=small}
\caption{Comparison of benchmarks with the manifold optimization algorithm of total throughput vs. number of MS1 columns $M_c$.} 
\vspace{-9pt}
\end{figure}

In Figs. 14 and 15, we compare our MIS-based design against three baselines: (i) a single static surface with our proposed algorithm “BCD-SCA, w/o MS2” and “manifold, w/o MS2” ), (ii) a sampling-assisted search with 2-bit and 3-bit phase quantization on the two MIS layers (grouped elements and uniform sampling when the configuration count exceeds $C_{\max}$), and (iii) a PSO-based heuristic that searches the continuous phase domain. For fairness, that is, the max-min user data rate or throughput, we instantiate our method with BCD-SCA on MIS; for total throughput, we use our manifold-based solver. All curves are evaluated under the same statistical channel settings. 

A brute-force global search over static phases and pattern scheduling is combinatorially intractable beyond toy sizes; therefore, we design a sample-assisted search-based scheme to find the best design within a large, quantized search space. Specifically, we quantize the static phases in both MIS layers 
\begin{align}
&\phi_m={e^{j2\pi b_\phi/2^{B_\phi}}}, b_\phi \in\mathcal{B}_\phi=\{1,\dots,2^{B_\phi}\},
\\
&\theta_n={e^{j2\pi b_\theta/2^{B_\theta}}}, b_\theta \in \mathcal{B}_\theta=\{1,\dots,2^{B_\theta}\}.
\end{align}
To manage complexity, we tie neighboring elements into contiguous groups of size $|\mathcal{G}_\phi|$ for MS1 and $|\mathcal{G}_\theta|$ for MS2, which yields an exhaustive enumeration over the grouped space with $C=2^{B_\phi\cdot \lceil \frac{M}{|\mathcal{G}_\phi|}\rceil}\times 2^{B_\theta\cdot \lceil \frac{N}{|\mathcal{G}_\theta|}\rceil}$combinations. When this count exceeds a cap $C_{\max}$, we uniformly sample $C_{\max}$ combinations for computational feasibility. This sampling-assisted search-based design retains the fidelity of an exhaustive search on medium scales and becomes a tractable proxy on larger scales. Within each evaluation, given $(\boldsymbol{\phi},\boldsymbol{\theta})$, we compute $\gamma_{k,u})$ for all users and beam patterns. Each user chooses $u_k^\star=\arg\max_{u}\log_2(1+\gamma_{k,u})$ using a greedy strategy, and finally, we evaluate the minimum user data or the total throughput to achieve the desired result.

Another benchmark is a heuristic algorithm based on particle swarm optimization (PSO). PSO is a stochastic swarm-intelligence optimization method that iteratively flies a group of particles in the search space, combining the information update speed and position of individual and global optima to approximate the global optimum. It does not require gradients, is simple to implement for solving continuous optimization problems that are non-convex, non-differentiable, or constrained by unit modules. The key steps of PSO include: \textbf{Particle encoding.} We optimize the phases in the angle domain: $\boldsymbol{\phi }=e^{j\boldsymbol{x}_{\phi}},\boldsymbol{\theta }=e^{j\boldsymbol{x}_{\theta}}$, and concatenate $\boldsymbol{x}=\left[ \boldsymbol{x}_{\phi};\boldsymbol{x}_{\theta} \right] \in [-\pi ,\pi) ^{M+N}$ when jointly optimizing both layers. \textbf{Fitness evaluation.} For a particle $\boldsymbol{x}:(\boldsymbol\phi,\boldsymbol\theta)$, the fitness functions are $f_{\min -\max}\left( \boldsymbol{x} \right) =\mathop {\max} \limits_{u\in \mathcal{U}}\sum_{k\in \mathcal{K}}{\log _2\left( 1+\gamma _{k,u}\left( \boldsymbol{x} \right) \right)}$ and $f_{\text{total}}\left( \boldsymbol{x} \right) =\mathop {\max} \limits_{u\in \mathcal{U}}\underset{k\in \mathcal{K}}{\min}\log _2\left( 1+\gamma _{k,u}\left( \boldsymbol{x} \right) \right)$ for minimum user data rate maximization and total throughput maximization, respectively. \textbf{PSO update.} For the $i$-th particle, in the $t$-th iteration, the position and velocity are $\boldsymbol{x}_{i}^{\left( t \right)}$ and $\boldsymbol{\nu }_{i}^{\left( t \right)}$, respectively. The individual and global optima are $\boldsymbol{p}_i$ and $\boldsymbol{g}$, respectively. The updating rules are
\begin{align}
&\boldsymbol{\nu }_{i}^{( t+1 )}\!=\!\omega \boldsymbol{\nu }_{i}^{( t )}\!+\!\varrho _1\boldsymbol{r}_1\odot ( \boldsymbol{p}_i\!-\!\boldsymbol{x}_{i}^{( t )} )\! +\!\varrho _2\boldsymbol{r}_2\odot ( \boldsymbol{g}\!-\!\boldsymbol{x}_{i}^{( t )} ),
\\
&\boldsymbol{x}_{i}^{\left( t+1 \right)}=\text{mod}( \boldsymbol{x}_{i}^{\left( t \right)}+\boldsymbol{\nu }_{i}^{\left( t+1 \right)}+\pi ,2\pi ) -\pi, 
\end{align}
where $\boldsymbol{r}_1,\boldsymbol{r}_2\sim \mathcal{U}\left[ 0,1 \right] ^{M+N}$ are random coefficient vectors with uniform distribution, $\omega$ are inertia, and $\varrho _1$ and $\varrho _2$ are cognitive and social coefficients, respectively. With the given swarm size, iterations, inertia, and cognitive and social coefficients, the PSO heuristically solves the original problems.

Across all settings, our MIS-based designs consistently outperform the benchmarks. Specifically, for the fairness objective, where $\kappa=10$dB, $M=6\times6$ and $N=4\times 4$ with varying numbers of users, the proposed BCD-SCA in MIS achieves the highest performance for all user loads because the BCD-SCA method constructs and exploits the convexity of the original problem for efficient continuous optimization. The PSO in MIS is trailing closely but remains below our method, while the sampling-assisted search lags further behind, improving from 2-bit to 3-bit quantization but still falling short of our method. The single-surface baseline is uniformly the worst, and the relative ordering is preserved as $K$ increases, collectively indicating that the benefits of differential position shift with static phases and coordinated pattern scheduling extend from light to heavy multiuser regimes. For total throughput versus $M_c$ with $\kappa=10$dB, $M_r=6$, $N=4\times4$ and $K=6$, although all methods grow steadily with aperture, the proposed manifold-based MIS solver yields the top curve throughout the range, maintaining a clear margin over PSO and both search variants. These results suggest that jointly and continuously optimizing the static phase shifts with beam pattern scheduling better exploits MIS degrees of freedom than stochastic PSO or discretized search, which are limited by parameter tuning and quantization, respectively.

\vspace{-12pt}
\subsection{Performance Comparison of the Proposed Algorithms}
\begin{figure}[t]
\centering
\includegraphics[width=2.8in,height=2in]{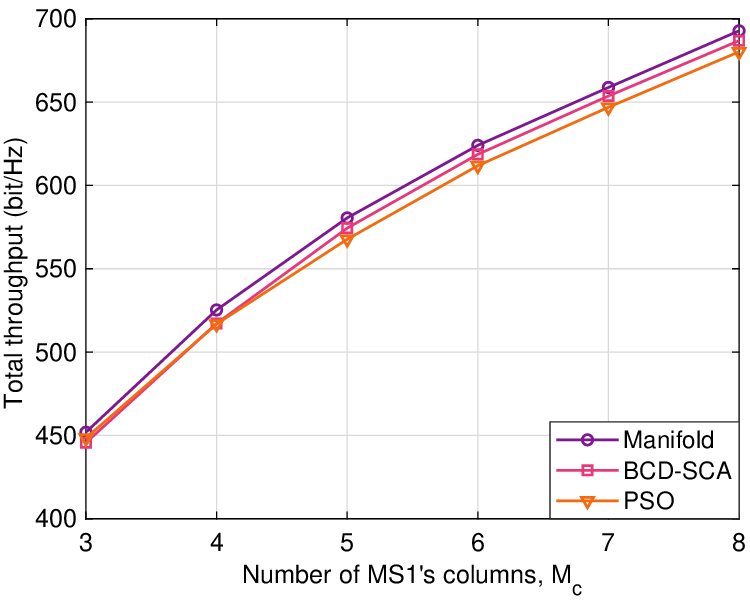}
\captionsetup{font=small}
\caption{Total throughput vs. number of MS1 columns $M_c$ of the algorithm based on manifold optimization, BCD-SCA, and PSO.} 
\vspace{-9pt}
\end{figure}
\begin{figure}[t]
\centering
\includegraphics[width=2.8in]{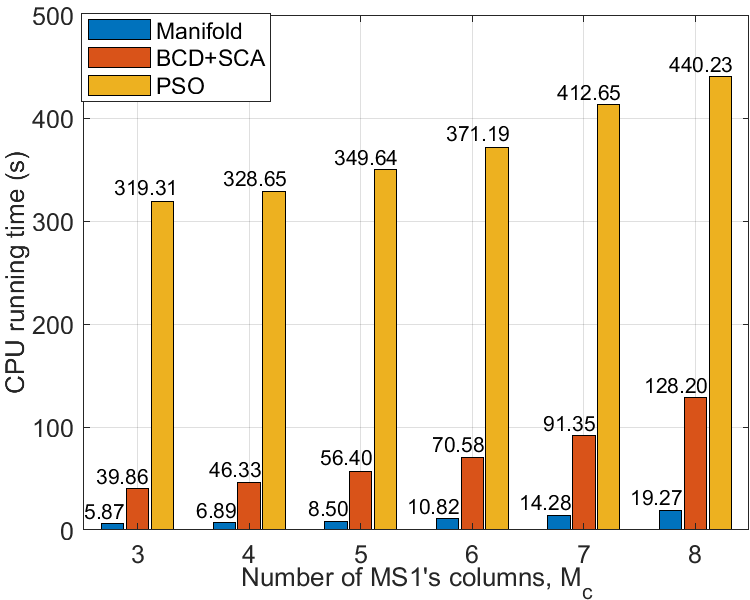}
\captionsetup{font=small}
\caption{CPU running time vs. number of MS1 columns $M_c$ of the algorithm based on manifold optimization, BCD-SCA, and PSO.} 
\vspace{-6pt}
\end{figure}

The penalty-assisted BCD-SCA Algorithm 1 is a more general optimization framework, which can also be applied to solve the total throughput maximization problem (P2), using the equivalent epigraph reformulation by extending the common slack variable for all users to a set of slack variables $\{\mu\}$ for each user. After that, similar procedures and algorithm framework in Section IV can be reused. When applying all three solvers to the same throughput-maximization problem (P2), Fig. 16 shows that all methods deliver near performance with a certain gap and scale almost linearly with $M_c$, with the manifold curve marginally on top, BCD-SCA a close second, and PSO slightly below both across all apertures. Fig. 17 reports CPU running time versus the number of MS1 columns $M_c$. The Riemannian manifold solver is consistently the most efficient, increasing from 5.87 s to 19.27 s as $M_c$ rises from 3 to 8. In contrast, the penalty-assisted BCD-SCA implementation requires 39.86 s to 128.20 s, while PSO takes significantly longer. Thus, the manifold method is roughly 6 times faster than the BCD-SCA framework and 23 times faster than PSO over this range, matching our complexity analysis $\mathcal{O}(I_{\mathrm{RCG}}KU(M+N)$ versus interior-point-dominated $\mathcal{O}(I_{\mathrm{in}}I_{\mathrm{out}}((KU)^3+M^3+N^3)))$. Hence, when (P2) is written as a smooth problem with only manifold feasibility, the Riemannian solver attains the best solution quality at a fraction of the runtime from the convex-approximation BCD-SCA scheme, while both decisively outperform a tuned PSO. 

\vspace{-12pt}
\subsection{Performance evaluation under CSI and phase error.}
\begin{figure}[t]
\centering
\includegraphics[width=3in, height=4.5in]{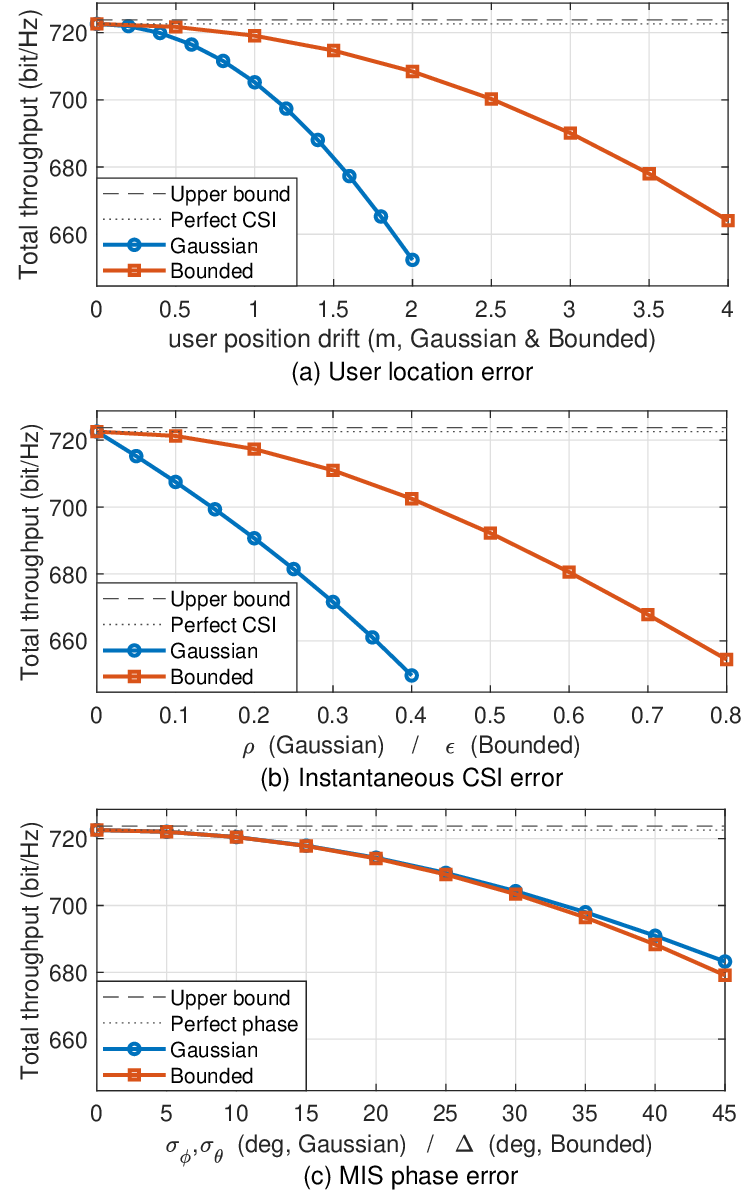}
\captionsetup{font=small}
\caption{Performance evaluation under CSI and phase error.} 
\vspace{-9pt}
\end{figure}

We optimize $(\boldsymbol{\phi},\boldsymbol{\theta},\boldsymbol{\Xi})$ under perfect CSI and then keep them fixed for robustness test. Three error sources are injected separately: (i) large-scale user location error, where the true user position is $\mathbf r_k^{\text{true}}=\mathbf r_k+\Delta\mathbf r_k$. We test a zero-mean Gaussian drift $\Delta\mathbf r_k\sim\mathcal N(\mathbf 0,\sigma_{\text{pos}}^2\mathbf I_3)$ and a bounded drift uniformly drawn in a 3D ball of radius $\varepsilon_{\text{pos}}$. Angles are recomputed from $\mathbf r_k^{\text{true}}$ while path-loss is kept fixed to avoid Jensen bias, thus isolating the impact of angular mismatch; (ii) Instantaneous CSI (small-scale) error, on the MIS–user link only (the BS–MIS channel is kept perfect): for the Gaussian case we use $\mathbf h_k^{\text{true}}=\sqrt{1-\rho}\mathbf h_k+\sqrt{\rho},\mathbf e_k$, where $\mathbf e_k\sim\mathcal{CN}\big(\mathbf 0,\tfrac{\alpha_{2,k}}{\beta_{2,k}+1}\mathbf S_{\text{mt}}\big)$, and for the bounded case we impose $\|\Delta\mathbf h_k\|_2\le\varepsilon_h\|\mathbf h_k\|_2$ with random directions; (iii) RIS phase error: in the bounded model both MS1 and MS2 phases are perturbed by per-element offsets $|\delta|\le\Delta$; in the Gaussian model only the movable MS2 has $\delta\sim\mathcal N(0,\sigma_{\theta}^2)$ accounting for positioning error. The figures report the perfect CSI and the upper bound as references.

All curves degrade monotonically with error, indicating a robust yet properly sensitive MIS design. For user location error, both Gaussian and bounded drifts cause gradual loss due to pointing mismatch, with a Gaussian drift of around 2 m reducing total throughput by 8\%, and a bounded radius of 4 m incurs 6-7\% loss. Instantaneous CSI error is the most detrimental per normalized mismatch; even a mild mismatch $\rho\approx0.2$ already causes a 4–5\% degradation. RIS phase error is comparatively benign in the tested range: Gaussian jitter on MS2 with $\sigma_\theta=30^\circ$ yields 2–3\% loss and $\sigma_\theta=45^\circ$ 5–6\%; bounded error on both layers shows a similar trend. These trends indicate that our differential position–based beam steering is intrinsically tolerant to moderate per-element phase jitter, and the optimized pattern scheduling preserves most of the MIS gain under realistic errors.

\vspace{-12pt}
\section{Conclusion}
In this paper, we proposed an MIS architecture that integrates the low complexity and cost advantages of static surfaces with specific dynamic beam steering capabilities of conventional RISs. 
By mechanically shifting a small, pre-phased MS2 over a larger, likewise static MS1, the MIS desired beam pattern switching without any electronic phase tuning, providing an attractive solution for quasi-static deployment scenarios. 
We built a binary-matrix model for MS2 positioning and formulated three mixed-integer non-convex optimization problems for MIS design under a TDMA protocol. To solve these, we developed a penalty-assisted BCD-SCA scheme for max-min fairness, an RCG-based algorithm on a product manifold for sum-rate maximization, and a penalty-assisted manifold optimization algorithm to quantify the performance ceiling under element-wise mobility of MS2. 
Comprehensive simulations demonstrate that the proposed MIS architecture substantially improves user rates and total throughput compared to single-layer static surfaces. Notably, even lightweight mechanical repositioning of a small MS layer yields considerable performance gains. The result also reveals that the moderate allocation of elements to MS2 maximizes the system utility, providing guidelines for hardware dimensioning. Furthermore, the element-wise MS2 mobility configuration can further recover up to 60\% of the performance gap between block-level MIS and fully dynamic RIS, underscoring the near-optimality of MIS with purely static phase profiles. 

Future research directions include hardware prototyping, the development of user mobility-aware MIS operation, the exploration of hybrid MIS designs that combine static and tunable elements, and the extension of the MIS concept to emerging advanced RIS architectures. Overall, MIS provides a practical and cost-effective pathway toward reconfigurable environments in next-generation wireless networks.

\ifCLASSOPTIONcaptionsoff
  \newpage
\fi



%\bibliographystyle{IEEEtran}

\end{document}